\documentclass[11pt]{article}

\usepackage{color}
\usepackage{amsmath}
\usepackage{mathrsfs}
\usepackage{amssymb}
\usepackage{amsthm}
\usepackage{epstopdf}
\usepackage{mathtools}
\usepackage[usenames,dvipsnames, table]{xcolor}
\usepackage[hang]{caption}
\usepackage{hyperref}
 \usepackage{bmpsize}

\usepackage{oldgerm}  
\usepackage[yyyymmdd,hhmmss]{datetime}

\usepackage{mathabx} 
\def\bee{\begin{enumerate}}\def\eee{\end{enumerate}}
\def\bei{\begin{itemize}}\def\eei{\end{itemize}}
\newcommand{\nco}{\newcommand}
\def\R{\mathbb{R}}
\nco{\red}{\color{red}}
\nco{\blue}{\color{blue}}
\nco{\brown}{\color{Magenta}}

\nco{\magenta}{\color{magenta}}

\nco{\violet}{\color{violet}}
\nco{\olive}{\color{Emerald}}
\nco{\orange}{\color{orange}}
\nco{\redend}{\normalcolor}
\nco{\blueend}{\normalcolor}
\def\inv#1{\frac{1}{#1}}
\def\tr{{\rm tr}\,}

\def\ommit#1{{}}
\def\({\left(}
\def\){\right)}
\def\ie{{\it i.e.,\/}\ }
\def\ie{{\rm i.e.,\/}\ }

\definecolor{cb}{rgb}{.8,.5,0}

\nco{\rnc}{\renewcommand}
\rnc{\title}[1]{{\Large\bf\mbox{}\\\medskip#1\bigskip\medskip\\}}
\rnc{\author}[1]{{\large #1\smallskip\\}}
\nco{\address}[1]{{\em #1\medskip\\}}

\def\diag{{\rm diag \,}}
\def\ii{\mathrm{i\,}}
\nco{\bun}{{\bf 1}}
\def\be{\begin{equation}}\def\ee{\end{equation}}
\def\bea{\begin{eqnarray}}\def\eea{\end{eqnarray}}
\def\bee{\begin{enumerate}}\def\eee{\end{enumerate}}
\def\bei{\begin{itemize}}\def\eei{\end{itemize}}
\def\oh{\frac{1}{2}}
\def\ommit#1{{}}

\def\SU{{\rm SU}}\def\U{{\rm U}}\def\SO{{\rm SO}}
\def\inv#1{\frac{1}{#1}}

\def\tr{{\rm tr\, }}

\def\eq=#1{\buildrel #1 \over{=}}

\def\CC{{\mathcal C}}
\def\CM{{\mathcal M}}
\def\CH{{\mathcal H}}     \def\CJ{{\mathcal J}}   \def\CO{{\mathcal O}}
\def\CP{{\mathcal P}}\def\CR{{\mathcal R}}\def\CS{{\mathcal S}}

\def\Ga{\pmb{\alpha}}

\def\c{c}  \def\c{r}    
\def\p{\mathrm{p}}

\def\gog{{\mathfrak g}}
\def\diag{{\rm diag \,}}
\def\ii{\mathrm{i\,}}
\def\N{\mathbb{N}}
\def\Z{\mathbb{Z}}

\def\R{\mathbb{R}}

\def\T{\mathbb{T}}

\def\hDelta{\hat\Delta}

\def\corr#1{{#1}}

\begin{document}
\begin{titlepage}
%
\begin{center}
\title{Revisiting Horn's Problem}
\medskip
\author{Robert Coquereaux} 
\address{Aix Marseille Univ, Universit\'e de Toulon, CNRS, CPT, Marseille, France}
\bigskip\medskip
\medskip
\author{Colin McSwiggen${}^{\dagger\ddagger}$ and Jean-Bernard Zuber${}^\ddagger{}^*$}
\address{
$\dagger$ Brown University, Division of Applied Mathematics, Providence, \corr{RI}, USA \\
 \ddag \ Sorbonne Universit\'e,  UMR 7589, LPTHE, F-75005,  Paris, France\\ \& CNRS, UMR 7589, LPTHE, F-75005, Paris, France\\
 \corr{* zuber@lpthe.jussieu.fr} 
 }

\bigskip\bigskip
{\sl Dedicated to the memory of Vladimir Rittenberg}
\bigskip\bigskip
\begin{abstract}
We review recent progress on Horn's problem, which asks for a description of the possible eigenspectra of the sum of two matrices with known eigenvalues. 

After revisiting the classical case, we consider several generalizations in which the space of matrices under study carries an action of a compact Lie group, and the goal is to describe an associated probability measure on the space of orbits. We review some recent results about the problem of computing the probability density via orbital integrals and about the locus of singularities of the density.  We discuss some relations with representation theory, combinatorics, pictographs and symmetric polynomials, and we also include some novel remarks in connection with Schur's problem.
\end{abstract}
\end{center}

 \end{titlepage}

 \section{Introduction}

 Horn's problem deals with the following question:  If  $A$ and $B$ are two $n$-by-$n$ Hermitian matrices with known eigenvalues 
$ \alpha_1 \ge \hdots \ge \alpha_n$ and $\beta_1 \ge \hdots \ge \beta_n\,$,  
what can be said about the eigenvalues $\gamma_1 \ge \hdots \ge \gamma_n$ of their sum $C = A+B$?   After almost a century of work,  starting with H. Weyl (1912) 
  and including  an essential conjecture by Horn \cite{Ho62}, 
 this problem is now solved, in the sense that there is a necessary and sufficient condition that is known to determine when $(\gamma_1,\cdots, \gamma_n)$ will occur as the spectrum of some such $C$ \cite{Kly, KT00}. In a nutshell, the $\gamma$'s must satisfy \corr{certain} linear inequalities, hence live inside a 
 convex polytope in $\R^{n-1}$.

 Horn's original problem may be generalized in different directions.
 First, Hermitian matrices, taken traceless and up to a factor $\ii$,  may be thought of as living in the Lie algebra of the group $\SU(n)$.
 One may consider as well the case of other simple Lie groups, and their so-called {\it coadjoint} orbits. Coadjoint orbits are known to carry the structure of a symplectic manifold and as such, Horn's problem has attracted the attention of symplectic geometers. 
 As we shall see, this coadjoint case may be treated in a detailed way.

 Another direction is to regard the Hermitian matrices as a case of {\it self-adjoint} complex $n$-by-$n$ matrices acted upon by the unitary group
 $\SU(n)$, and to consider in parallel the two other 
 cases of real symmetric matrices under the action of the real orthogonal group $\SO(n)$, and of quaternionic self-dual matrices under
 the action of the unitary symplectic group $\mathrm{USp}(n)$.

 In this paper we review some  recent progress regarding both of these generalizations.  First, we discuss what is known regarding the similarities and differences between the three self-adjoint cases, and we explain the relationship between Horn's problem and Schur's problem of characterizing the possible diagonal entries of a matrix with known eigenvalues. Second, the study of coadjoint orbits 
 leads to very interesting connections with representation theory and combinatorics, namely the determination of multiplicities 
 in the decomposition of tensor products of irreducible representations.   It has been known for a while that Horn's problem 
 yields a {\it semi-classical} approximation to that problem, in the limit of large representations  (\ie tensor products of irreps whose highest weights lie deep in the dominant Weyl chamber). We present below an alternative approach, through an exact relationship between the two problems  that is more precise than the previously known asymptotic relationship. 
 %
 
 The questions addressed in this article are primarily of mathematical interest, and some of the cases that we study are not yet known to have a direct physical application.  However, many mathematical objects that arise in this investigation, such as tensor product multiplicities and orbital integrals, are ubiquitous in today's theoretical physics, and an improved understanding of these objects represents an expansion of the future physicist's toolkit. The possibility to extend the considerations of this paper to the current (affine) algebras and to the fusion of their representations is an especially interesting route to explore.  At any rate, we hope that with its many facets in many directions, this subject would have pleased our colleague and friend Vladimir, who was a man of culture and of tireless curiosity.

\section{What is Horn's problem ?}
 \subsection{Introduction}
 Given two Hermitian $n\times n$ matrices $A$ and $B$, of known spectrum 
 $$ \corr{\alpha_1 \ge \hdots \ge \alpha_n\qquad \mathrm{and}\qquad \beta_1 \ge \hdots \ge \beta_n\,,}$$ 
 what can be said about the spectrum
  ${\gamma=\{\gamma_1 \ge \gamma_2\ge \cdots \ge \gamma_n\}}$  of their sum ${ C=A+B}$ ?
 
 This is an old problem, with a rich history~\cite{DST, Fu}. 
 Obviously, {$\sum_{k=1}^n \gamma_k - \alpha_k - \beta_k=0$}, thus $\gamma$ must lie in an $(n-1)$-dimensional hyperplane that we identify with
 $\R^{n-1}$.
 It is clear that $\gamma$ must additionally satisfy some linear inequalities to belong to the spectrum of $A+B$. 
 For example, we have the obvious inequality {$\gamma_1 \le \alpha_1+\beta_1$} 
 stemming from  the  maximum principle $\gamma_1=\max_\psi \frac{\langle \psi , (A+B)\psi\rangle}{\langle \psi,\psi\rangle}$,  
 or {Weyl}'s inequality~\cite{We}, 
 $1\le i,\ j,\ i+j-1\le n\ \Rightarrow \ {\gamma_{i+j-1}\le \alpha_i+\beta_j }$. 
{Horn}~\cite{Ho62} conjectured
 a set of {\it  linear} inequalities relating $\alpha$, $\beta$ and $\gamma$ which are {\it necessary and sufficient} 
 conditions for $\gamma$ to occur as the spectrum of some $A+B$:
 $${ \sum_{k\in K} \gamma_k \le \sum_{i\in I} \alpha_i +\sum_{j\in J} \beta_j }$$
 for some triplets $\{I,J,K\}$  of subsets of $\{1,\cdots, n\}$ of the same cardinality,  which may be determined recursively. 
 
 If true (and it is!), Horn's conjecture implies that the possible values of $\gamma$ lie in a {\it convex polytope} in $\R^{n-1}$. This convexity property is not a surprise in the context of symplectic geometry (Atiyah--Guillemin--Sternberg--Kirwan convexity theorems)~\cite{Knut}.

After many contributions by many mathematicians, Horn's conjecture was finally proven by {Klyachko} \cite{Kly} 
and {Knutson and Tao}~\cite{KT99}. 

The problem is interesting for its many facets and ramifications, its interpretation in symplectic geometry, 
its appearance in various guises -- algebraic geometry (via a connection with Schubert calculus), invariant factors, among others -- and its connections with representation theory and combinatorics. We refer the reader to the review by Fulton~\cite{Fu}.

\subsection{The classical Horn problem revisited}

Horn's  problem may be reformulated  as follows.
Let $\CO_\alpha$ be the {\it orbit} of $\diag(\alpha_1, \alpha_2,\cdots,\alpha_n)$
under the adjoint action of $\U(n)$, 
$$ { \CO_\alpha=\{\, U\diag(\alpha_1, \alpha_2,\cdots,\alpha_n) \corr{U^{-1}}\ |\ U\in \U(n)\, \}}$$
and likewise $\CO_\beta$.  Then which orbits $\CO_\gamma$ intersect the (Minkowski) sum of orbits  $\{A + B \, : \, A \in \CO_\alpha, B \in \CO_\beta \}$? 

(i) In particular, suppose we take $A$ {\it uniformly distributed} on $\CO_\alpha$ (according to the Haar measure), 
and likewise $B$ uniform on $\CO_\beta$ and {\it independent} of $A$.
Can one determine the probability distribution of~$\gamma$, i.e. by explicitly writing down its probability density function (PDF)?

(ii) Traceless Hermitian $n$-by-$n$ matrices may be regarded as  elements of the dual of the Lie algebra of $\SU(n)$.  The action of SU(n) on these matrices by conjugation is its coadjoint representation. What happens if we  consider sums of orbits in the coadjoint representations of other classical Lie groups? Can we compute the probability distribution as in (i)?

(iii) Finally, what happens if we replace orbits of complex Hermitian matrices under the conjugation action of  $\U(n)$ by

-- orbits of real symmetric matrices under the  conjugation action of { $\SO(n)$}, or

-- orbits of quaternionic Hermitian (aka self-dual) matrices under the conjugation action of $\mathrm{USp}(n)$ ?

In all three cases,
any matrix $A$ may be brought to a diagonal form $\diag(\alpha_1,\ldots, \alpha_n)$ by conjugation by some element of the group. 

For questions (i) and (ii), as we'll see below, the answer is Yes, we can!  For the last question, much less is known.
There is, however, a general result by {Fulton}, which asserts that 
 Horn's inequalities relating $\alpha,\beta, \gamma$ are the same in the three ``self-adjoint" cases. Hence, for given $\alpha$ and $\beta$, the set of possible $\gamma$ is the 
same polytope irrespective of the class of self-adjoint matrices. 
What about the distribution of $\gamma$, if again $A$ and $B$ are uniformly and independently distributed on their orbits?

\begin{figure}[!tbp]
  \begin{center}
       \includegraphics[width=12pc]{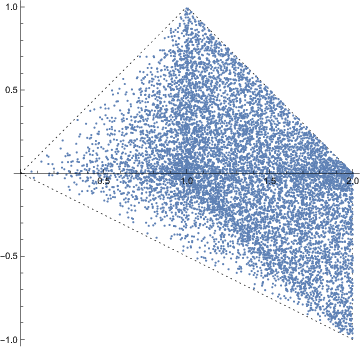}\quad   \includegraphics[width=12pc]{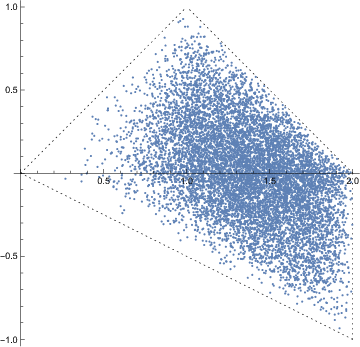}   \quad\includegraphics[width=12pc]{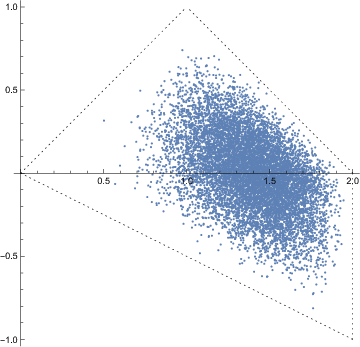}   \\
         \includegraphics[width=12pc]{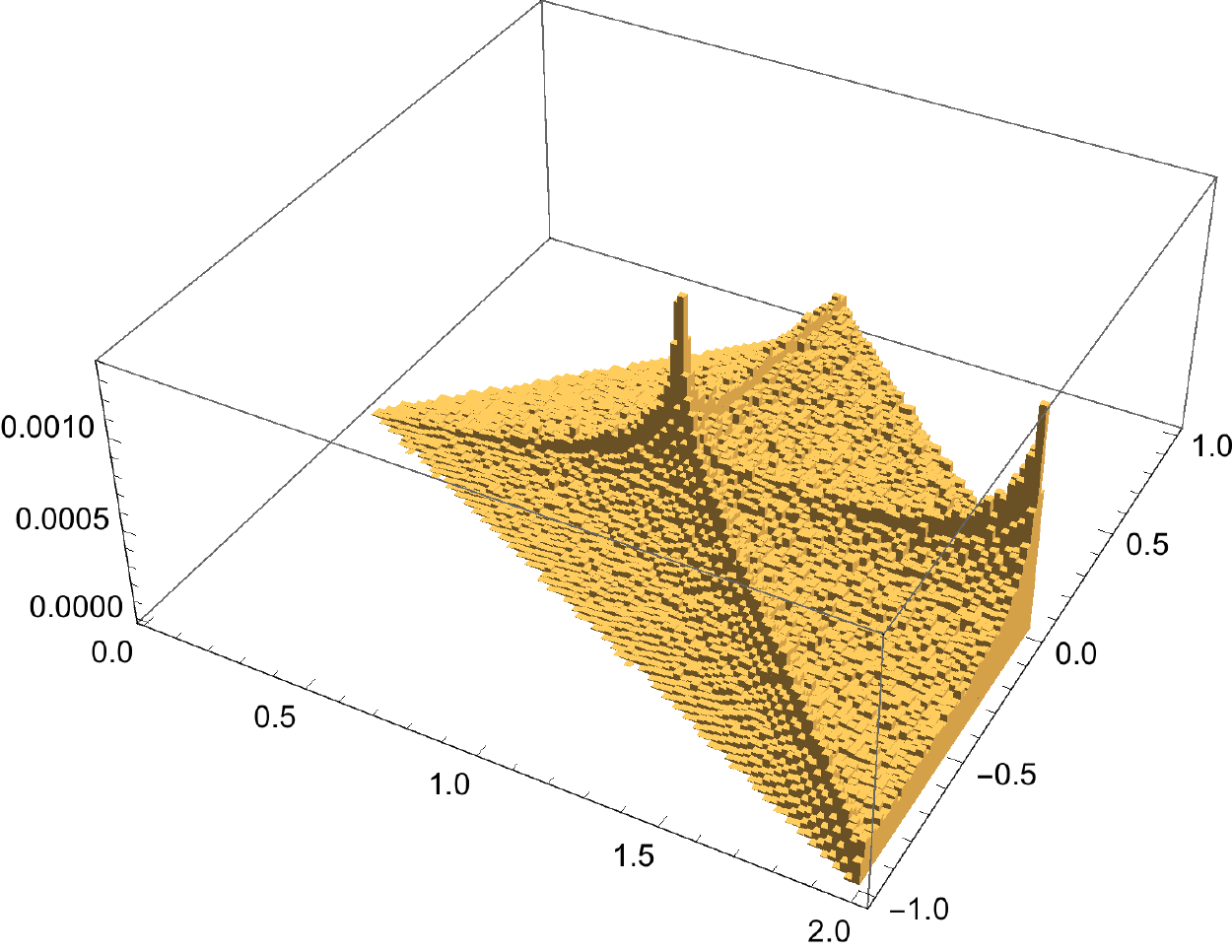} \quad  \includegraphics[width=12pc]{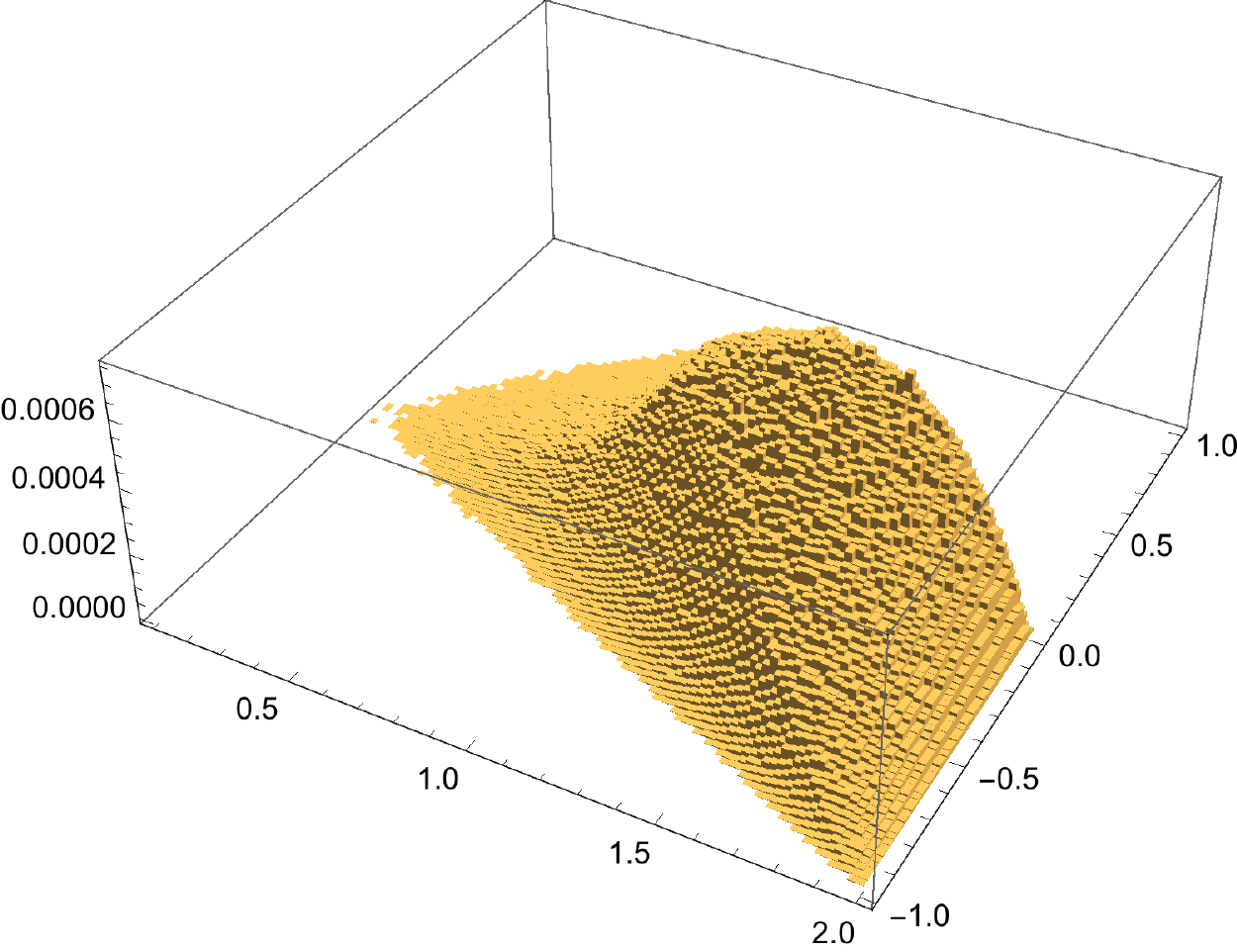}\quad      \includegraphics[width=12pc]{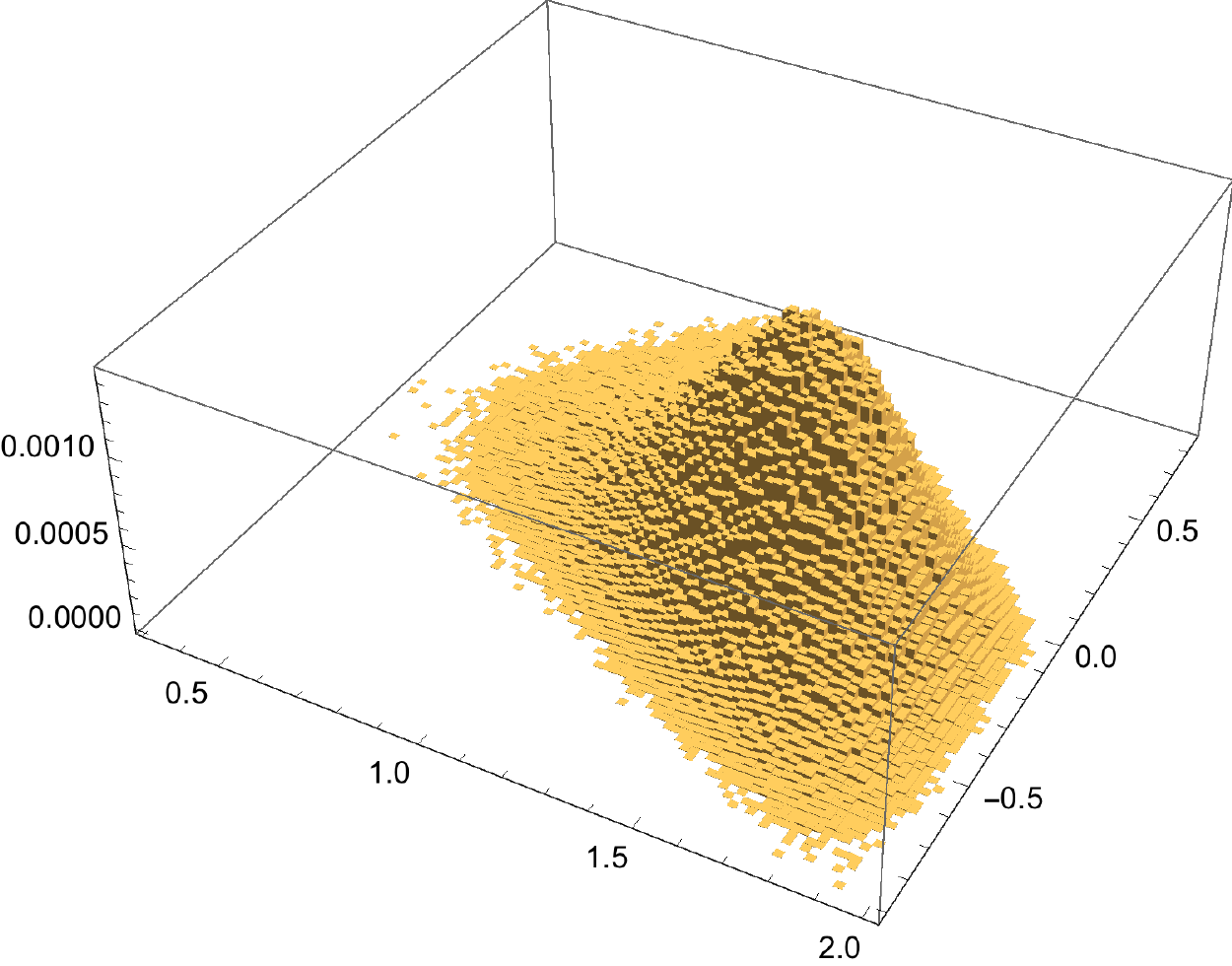}\\[10pt]
    \caption{\small Comparing the action of  SO(3) (left), SU(3) (middle) and USp(3) (right)  on $\alpha=\beta=(1,0,-1)$. }
        \label{comparison}
      \end{center} \end{figure}

It is revealing to make a (numerical) experiment. 
Take for example $n=3$ and $\alpha=\beta=(1,0,-1)$, and generate using Mathematica \cite{Mathematica} many samples of 
$C=\diag(\alpha) +  V\diag(\beta) V^{-1}$, with $V$ drawn randomly from the Haar measure on the appropriate group.  
 Diagonalize the samples and plot 
$(\gamma_1 , \gamma_2)$. Recall that by convention $\gamma_1\ge \gamma_2\ge \gamma_3=-\gamma_1-\gamma_2$. See Fig. \ref{comparison}.

 We observe, as expected, that $(\gamma_1,\gamma_2)$ lies in the same convex polygon in the three cases.
 The fact that the
    distribution is more  concentrated about its mean as we go  from SO to SU to USp   
    is also natural, as a consequence of a Jacobian prefactor in the 
    PDF, see below. But the most  striking (and unexpected) feature is the appearance of 
       lines of enhancement in the SO(3) case. These lines become even more conspicuous when one computes the histogram
       of the three distributions.

It should also be stressed that these features do not depend on the particular choice of $\alpha$ and $\beta$ that we have made here.
See \cite{Z1} for other examples exhibiting the same singularities.

{\it Question}:
 Can one compute the PDF for the three cases and understand the origin, location and nature of the singularities
 in the orthogonal case?
 
 We shall see below that the PDF for the unitary and symplectic cases admits a closed-form expression, whereas an explicit expression in the orthogonal case appears out of reach.  Nonetheless, we can determine a great deal about the singularities of the PDF in all three cases.

\subsection{From Schur to Horn}
Before we proceed, let us first observe that there is a limiting case of Horn's problem where it reduces to another well-studied problem, namely Schur's problem:

{\sl Given a matrix $A$ on the orbit $\CO_\alpha$ (of any of the types previously discussed), 
what can be said about the diagonal matrix elements of $A$?}

It is known \cite{Ho54} that the diagonal matrix elements of $A$ lie in the {\it permutahedron} $\CP_\alpha$, \ie the convex
polytope  with vertices $(\alpha_{P(1)},\ldots,\alpha_{P(n)})$, $P\in \CS_n$. 
More precisely, if $A$ is drawn uniformly at random from its orbit $\CO_\alpha$, what is the distribution of the diagonal elements $\xi_i:=A_{ii}$?
For $\SU(n)$ orbits, it is known that this distribution coincides with the (normalized) {\it Heckman measure}~\cite{He82}, whose density is a piecewise polynomial function of degree $(n-1)(n-2)/2$. 
 For $\SO(n)$ orbits,
much less is known~\cite{Fa}.


\ommit{  \begin{figure}
\begin{center}
\includegraphics[width=0.4\textwidth]{Schur-determinations-a.pdf}
\caption{Piecewise polynomial determination of the PDF of the $\xi_i:=A_{ii}$, for SU(3) }  
\label{Schur-determinations}
\end{center}
\end{figure}}

\begin{figure}
\begin{center}
\includegraphics[width=0.4\textwidth]{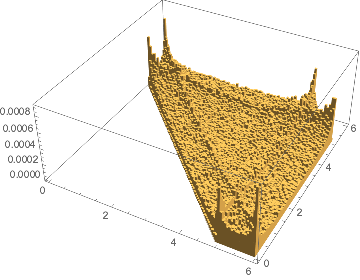}
\caption{Histogram of the diagonal elements $(A_{11},A_{22})$, for $A=O \diag(7,4,-11) O^{-1} $, $O\in \SO(3)$.}  
\label{Schur-histo-650}
\end{center}
\end{figure}

Take first the case $n=2$ and a traceless diagonal matrix $\diag(\alpha_1,-\alpha_1)$. For SO(2), resp. SU(2), orbits, the PDF of the $A_{11}$ element is readily computed
\be\label{Schur-n=2} \p(A_{11})=\begin{cases} \begin{cases}\frac{ \mathrm{const.}}{\sqrt{\alpha_1^2-A_{11}^2}} & \SO(2)\\
 \mathrm{const.} &\SU(2) \end{cases}  \quad 
&\mathrm{for}\ 0\le |A_{11}|\le |\alpha_1| \\ 0 & \mathrm{otherwise}.\end{cases} \ee
Thus we find that the PDF exhibits an integrable inverse--square-root edge singularity in the $\SO(2)$ case. For SO(3) orbits, numerical experiments show  a singular behavior, see Fig.~\ref{Schur-histo-650}.  
 
\begin{figure}
\begin{center}
\includegraphics[width=0.2\textwidth]{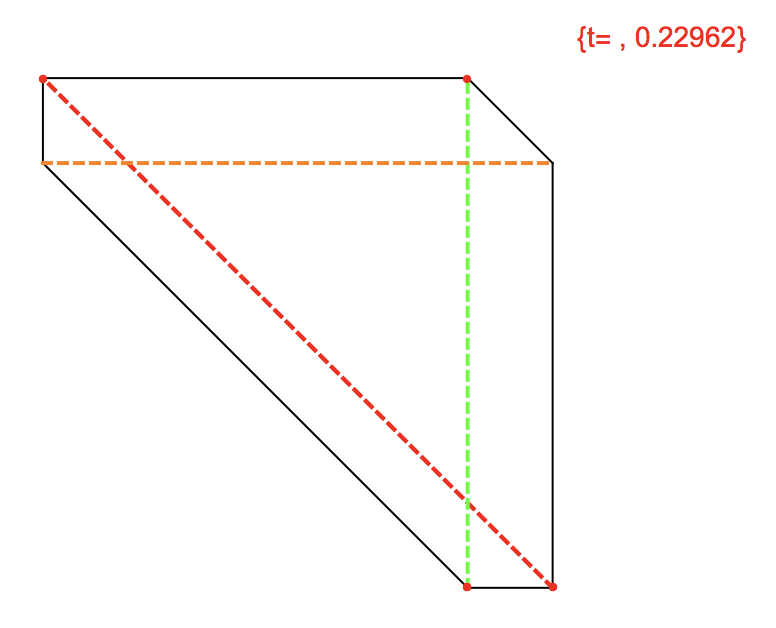}\includegraphics[width=0.2\textwidth]{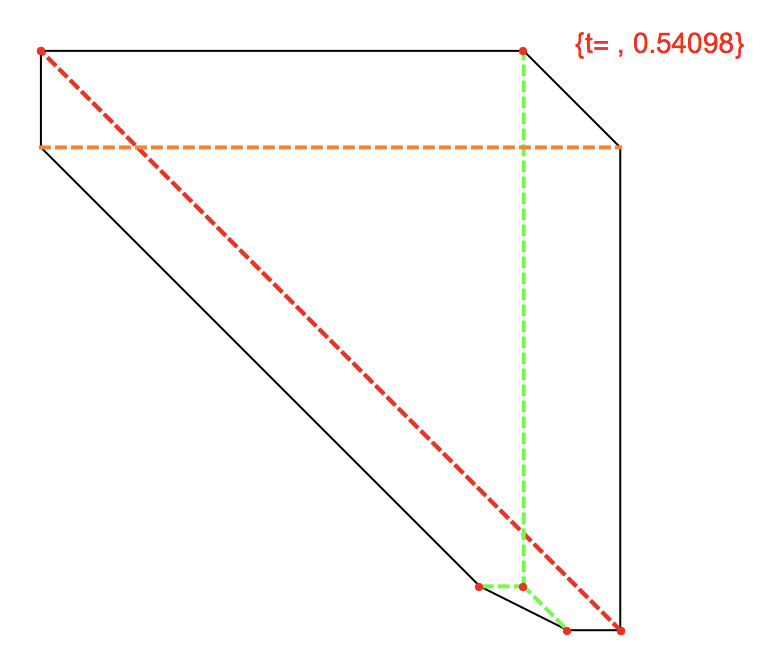}\includegraphics[width=0.2\textwidth]{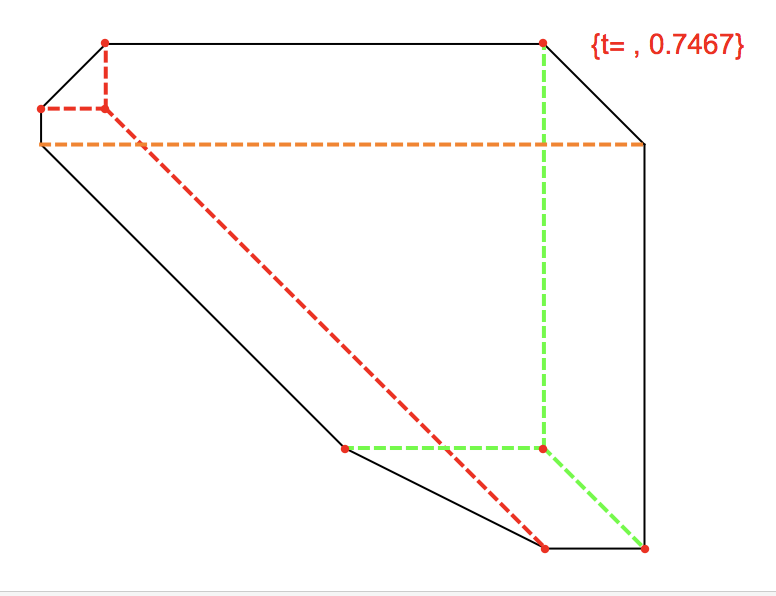}\includegraphics[width=0.2\textwidth]{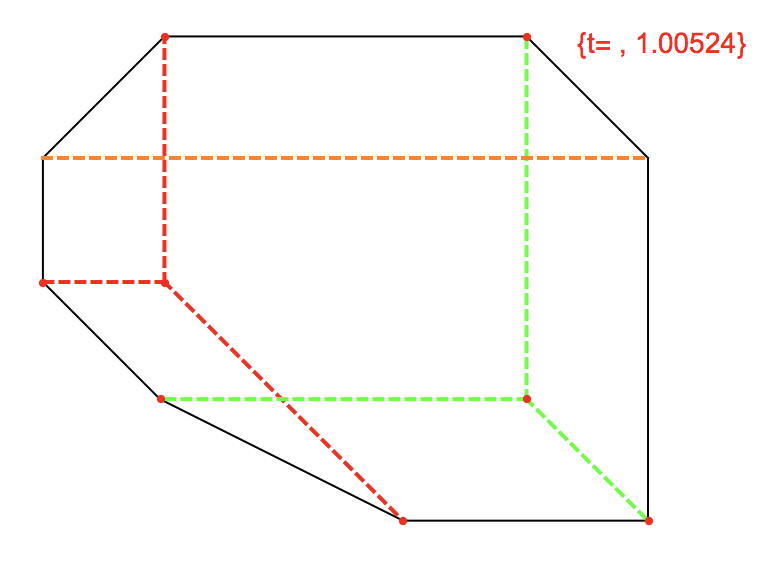}\includegraphics[width=0.2\textwidth]{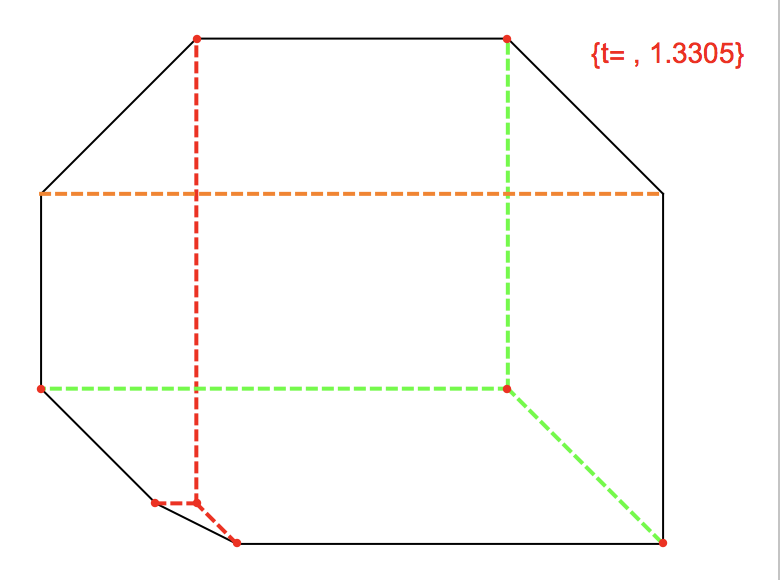}
\caption{Interpolating polygon and singular lines from Schur to Horn, for $ t g (7,4,-11) g^{-1} + (11,-1,-10) $.
For $0\le t\le \oh$, the singular lines are the diagonals of the (shifted) permutahedron; for  $\oh\le t\le \frac{2}{3}$, a triple 
point appears along with a new side of the polygon, for $2/3\le t\le 7/4$, a second triple point appears etc. }
\label{schur2horn}
\end{center}
\end{figure}

There is an obvious relationship with Horn's problem. Given two 
matrices $A\in \CO_\alpha$ and $B=\diag(\beta)$, Horn's problem for $C=t A+B$, $t$ small, reduces to Schur's problem. Indeed to first order in perturbation theory in $t$, the eigenvalues $\gamma(t)$ of $C$ are
$$ \gamma_i(t)= \beta_i + t A_{ii} +O(t^2)\,.$$
Thus to first order, the 
Horn polytope is nothing but the permutahedron $\CP_\alpha$ shifted by the vector $\beta$. 
It is interesting to see how the polygon of support and the singular lines deform
as $t$ grows, see Fig.~\ref{schur2horn} and the discussion in the next section.

\subsection{The locus of  singularities in Horn's problem}
  \label{singlocus}
 Compare the three ``self-adjoint cases'' of real symmetric, complex Hermitian or  quaternionic self-dual, $n\times n$ (traceless) matrices. We label these cases by a parameter ${\theta}=\oh,\, 1,\, 2$, i.e. {\it half} the Dyson index familiar from random matrix theory, as follows:

\begin{center}\begin{tabular}{|c| c|c|}
\hline $\theta$ &$\CM^\theta_n$ & $G_{\theta}$ \\
\hline
$\oh$ & Real Symmetric & $\SO(n)$ \\
\hline
1 & Complex Hermitian & $\SU(n)$\\
\hline
2 & Quaternionic Self-Dual & $\mathrm{USp}(n)$\\
\hline
\end{tabular}
\end{center}

 For given $n$ and $\alpha, \beta$, not only the support of the distribution of $\gamma$ is the same \cite{Fu}, 
 but also the {\it locus} of singularity of the PDF (although the singularities themselves are of quite a different nature). We may state the following proposition~\cite{CMSZ}:\\
 
 {\sl The PDF is a piecewise real-analytic function of $\gamma$. Non-analyticities occur only when $\gamma$ lies on hyperplanes of the form
\be\label{non-anal} \sum_{k\in K}\gamma_k = \sum_{i\in I} \alpha_i +\sum_{j\in J}  \beta_j\ee
 with $I,J,K\subset \{1,\cdots,n\}, \ |I|=|J|=|K|$, independently of $\theta$.}\\
 
 \textbf{Hint at proof:} Consider the map $\Phi$ : $G\times G\to \CM^\theta_n$, $(g_1,g_2) \mapsto C=A+B= g_1 \alpha g_1^{-1} +g_2 \beta g_2^{-1}$.  If $C$ is a regular value of $\Phi$ in the sense that the differential $d\Phi$ is surjective at all points of the preimage $\Phi^{-1}(C)$, then the PDF is real analytic at $\gamma$. Non-analyticities can therefore only occur at $\gamma$ that are spectra of non-regular values of $\Phi$, and it is easy to see that these lie on hyperplanes of the form (\ref{non-anal}), see details in \cite{CMSZ}. \\

\textbf{Remarks:}\\ -- This condition encompasses boundary  facets of Horn's domain other than  those lying on the hyperplanes $\gamma_{i}=\gamma_{i+i}$, where indeed the PDF vanishes in a non-analytic way. \\
-- Eq. (\ref{non-anal}) gives a necessary, but not sufficient, condition for non-analyticity. It doesn't say where singularities do in fact occur. Typically, 
we'll find that singularities appear only on subsets of the hyperplanes defined by (\ref{non-anal}).

 \section{Computing the PDF}
  \subsection{The orbital integrals}
  \label{sect-orb-int}
 
As will appear, in all cases 
a central role is played by the  
{\it orbital integral} { (aka generalized or multivariate Bessel function)}.
In the self-adjoint cases, we write
$$ \CH_{\theta}(A, X) =  \int_{G_{\theta}} \exp(\tr (X  g A g^{-1})) \, dg$$
where $A, X \in \CM_n^\theta$ and $dg$ is the normalized Haar measure.  In the coadjoint cases, we rather write it as 
\be\CH_\gog (A, X) =  \int_{G} \exp( \langle X , \mathrm{Ad}_g A \rangle)) \, dg\ee
where $A,X\in \gog$, the Lie algebra of $G$, and $\langle \cdot, \cdot \rangle$ is a $G$-invariant inner product. 
\corr{When the case under consideration is clear, we shall suppress the subscript and write both types of integral as $ \CH(A, X)$.}

Note that:

$\bullet$ As a function of $X$, $\CH_{\corr{\gog}}(A,iX)$  is the Fourier transform of the {\it orbital measure} at $A$, i.e. the unique $G$-invariant probability measure concentrated on the orbit of $A$. 

$\bullet$ $\CH_{\theta}(A,X)$ depends only on the eigenvalues $\alpha$ and $x$ of $A$ and $X$. 
Likewise, $\CH_{\gog}(A,X)$ depends only on $\alpha$ and $x$, representatives of the orbits of $A$ and $X$ in 
the dominant chamber of a Cartan subalgebra $\mathfrak{t}$.
With a small abuse of notation, 
\corr{we shall often write the integral as}  $\CH(\alpha,x)$.
   
In the unitary ($\theta=1$) case with Hermitian (or anti-Hermitian) matrices, 
the explicit formula is well known to physicists  under the HCIZ acronym~\cite{HC,IZ}. 
For $A$ and $X$ ``regular,'' i.e. $\alpha_i\ne \alpha_j$ and $x_i\ne x_j$, 
$${ \CH_{1}(\alpha, \ii x) }
={ 
\int_{\SU(n)} e^{\ii \tr(X  V A V^*)} \, dV}
={ \prod_{p=1}^{n-1} p! 
\, \frac{\mathrm{det} \corr{(} e^ {\ii x_i \alpha_j})_{1 \leq i, j \leq n}}{\Delta(ix) \Delta(\alpha)}}\,,$$
 where $\Delta(x)=\Pi_{i<j} (x_i-x_j)$ is the Vandermonde determinant of the $x$'s.

This is a particular case of a general formula due to Harish-Chandra, \corr{\cite{HC}, see also \cite{McS}}
\be\label{HC-form}\CH_{\gog}(x)= \mathrm{const.}\, \sum_{w\in W} \epsilon(w)\frac{e^{\ii \langle w(\alpha), x\rangle}}{\Delta_\gog(\ii x) \Delta_\gog(\alpha)}\,.\ee
Here $W$ is the Weyl group, $\epsilon(w)$ is the signature of $w\in W$, and 
\be\label{Deltag}\Delta_\gog(x):=\prod_{\Ga>0} \langle \Ga,x\rangle\,,\ee
a product over the positive roots\footnote{The reader should not confuse the roots $\Ga$ of the algebra $\gog$ with the eigenvalues $\alpha$ of Horn's problem.}, generalizes the Vandermonde determinant to all $\gog$, see \cite{McS}. 

Returning to the self-adjoint cases, in the symplectic ($\theta=2$) case, there is a generalization of
the previous formulae  which reads~\cite{BH}  
$${ \CH_{2}(\alpha, \ii x) ={\rm const.}
\, \sum_{P\in \CS_n} \, \frac{ e^ {\ii \sum_j x_j \alpha_{Pj} } }{\Delta^3(ix) \Delta^3(\alpha_P)}} {f_n(x,\alpha_P)}\,,$$
where $f_n$ is a polynomial in the variables $\tau_{i,j}:= (x_i-x_j) (\alpha_{Pi}-\alpha_{Pj})$, $\deg(f_2)= 1$, $\deg(f_3)= 3$, etc.
A recursive formula is known to construct $f_n$ for higher $n$. 

In the orthogonal ($\theta=\oh$) case, most unfortunately, there is no similar compact expression. The best that may be achieved 
is a series expansion in terms of zonal polynomials (see \cite{CZ2} and references therein), which is not very handy for 
detailed calculations.

 \subsection{The Horn PDF in terms of orbital integrals}
 We may now state the following integral representation of the Horn PDF.  Here we assume again that $A$ and $B$ are traceless, and we also assume they each have $n$ distinct eigenvalues.\footnote{These assumptions are easily removed, but without them the ``PDF'' may additionally include a delta distribution enforcing the constraint $\mathrm{tr}(C) = \mathrm{tr}(A) + \mathrm{tr}(B)$, as well as other linear constraints due to degenerate eigenvalues.}  We identify the space of $n$-by-$n$ traceless diagonal matrices with $\R^{n-1}$.  Then we have:

 {\sl For  self-adjoint matrices $A$ and $B$, independently and uniformly distributed on their $G_{\theta}$-orbits $\CO_\alpha$ and $\CO_\beta$,
the PDF of $\gamma$ is given by
\be\label{PDF} \p(\gamma\vert\alpha, \beta) =  
 {\rm const}{({\theta},n)} 
 \,   |\Delta(\gamma)|^{\theta} \int_{\R^{n-1}} d^nx \, |\Delta(x)|^{2\theta} \, {\CH_{\theta}}(\alpha, \ii x) {\CH_{\theta}}(\beta, \ii x) {\CH_{\theta}}(\gamma, \ii x)^* \,,\ee
where $\Delta(x)=\prod_{i<j}(x_i-x_j)$ is the  Vandermonde determinant.

For coadjoint orbits, a similar formula applies, where the integral runs over the Cartan subalgebra $\mathfrak{t}$ and $|\Delta(x)|^{2\theta}$ is replaced by $\Delta^2_\gog(x)$. 
}

The proof is elementary:  $\CH(\alpha,\ii x)$, the Fourier transform of the orbital measure, may also be regarded as the characteristic function 
(in the sense of probability theory) of the random variable $A\in \CO_\alpha$. The characteristic function of $C=A+B$ is the 
product $\CH(\alpha, \ii x)\CH(\beta,\ii x)$, from which the PDF of $C$ is recovered by inverse Fourier transform. As the latter depends 
only on the eigenvalues $\gamma$, paying due care to  the Jacobians that occur in the changes from matrices to eigenvalues, one finds 
(\ref{PDF}). 
\ommit{1. 
 For a random variable $A$ uniformly distributed on its orbit $\CO_\alpha$, the  characteristic function is
 $${\varphi_A(X) :=\mathbb{E}(e^{\ii \tr X A}) 
=\int_{G_{\theta}} DV \exp(\ii \tr  X V{\alpha}V^*)= \CH_{\theta}(\alpha, \ii x)}$$
2. Since $A$ and $B$ are independent, characteristic function of
the sum $C=A+B$ is the product
$${\varphi_C(X)= \mathbb{E}(e^{\ii \tr X C})=  \varphi_A(X) \varphi_B(X)}\,. $$\\[2pt]
3. The PDF of $C$ then recovered by  inverse Fourier transform 
$$ {\p(C|\alpha,\beta)=\inv{(2\pi)^{\#}} \int DX e^{-\ii \tr X C}  \varphi_A(X) \varphi_B(X)\,.}  $$
4. $\varphi_A(X)$ depends only of eigenvalues $x$ of $X$, and $\p(C|\alpha,\beta)$ only on 
eigenvalues $\gamma$ of $C$, hence define PDF of the $\gamma$'s. Don't forget the Jacobians !
$$ DX ={\rm const.} |\Delta(x)|^{\theta} \prod_{1\le i\le n} dx_i \qquad \qquad  \Delta(x)=\prod_{i<j}(x_i-x_j) \quad {\rm Vandermonde\ det}$$
\bea {\p(\gamma|\alpha,\beta)}&=&
{\rm const.}{ |\Delta(\gamma)|^{\theta}} \, { \p(C|\alpha,\beta) }
\\
  &=& {   {\rm const. } 
    |\Delta(\gamma)|^{\theta} \int_{\R^n} \prod_{i=1}^n dx_i\, |\Delta(x)|^{\theta}\,  \CH_{\theta}(\alpha,\ii x)\CH_{\theta}(\beta,\ii x)
  \CH_{\theta}(\gamma,\ii x)^*\,.} \qquad \Box \eea}

  This formula must have been known to a number of people, 
  see in particular \cite{Dooley-etal} and other related references in \cite{Z1}.

 \subsection{Explicit computation of the PDF $\p(\gamma)$ in the  $\SU(n)$ case.}
It is then a matter of simple calculation to write an explicit form of the PDF for low values of $n$.
 One finds that the PDF is the product of a normalizing constant, a ratio of Vandermonde determinants and a function that we call $\CJ$, which will soon gain a geometrical interpretation:
\bea\label{p-Jn}  \p(\gamma|\alpha,\beta)&=& \frac{\prod_1^{n-1} p!}{ n!}  
\frac{ \Delta(\gamma)}{\Delta(\alpha)\Delta(\beta)}  {\CJ(\alpha,\beta;\gamma) }
\\ \label{Jn}
\CJ(\alpha,\beta;\gamma)
 &=&\,    \frac{\ii^{-n(n-1)/2}}{(2\pi)^{n-1}}   
 \sum_{P,P'\in S_n}\varepsilon_P\,\varepsilon_{P'}\, 
\int \frac{d^{n-1}u}{\widetilde\Delta(u)}\,\prod_{j=1}^{n-1}  e^{\ii u_j A_j(P,P',I)}\,,\eea
 where $A_j(P,P',P'')= \sum_{k=1}^j (\alpha_{P(k)}+\beta_{P'(k)}-\gamma_{P''(k)}) 
$  and {$\widetilde\Delta(u):=\prod_{1\le i<j\le n}(u_i+\cdots+u_{j-1})$}.

Note that $\CJ$ is a linear combination of integrals over $u\in \R^{n-1}$ of the form $\int \frac{d^{n-1}u}{\widetilde\Delta(u)}  e^{\ii u_j A_j}$,
generalizing the classical Dirichlet integral
$$ {\mathcal{P}} \int_{\R} \frac{du}{u} e^{\ii u A} = \ii \pi \epsilon(A)\,, \qquad \mathrm{if}\ A\ne 0,$$
where $\epsilon$ is the sign function and ${\mathcal{P}}$ is Cauchy's principal value. Carrying out a partial fraction decomposition of $1/\widetilde\Delta(u)$ into  simple elements  and using repeatedly
$${\mathcal{P}}\int_\Bbb{R} \frac{du}{u^r} e^{\ii u A} =\ii \pi \frac{(iA)^{r-1}}{(r-1)!} \epsilon(A)\,$$
leads to very explicit expressions for $\CJ$. This has been carried out for $n=2,\cdots,6$ in \corr{\cite{Z1,CZ1}}. \\

\noindent -- $\CJ$ is clearly a homogeneous function of $(\alpha,\beta,\gamma)$ of degree $(n-1)(n-2)/2$.\\
-- It is  discontinuous for $n=2$, where it is, in the variable $\gamma_1-\gamma_2$, \
just the indicator function of the interval $[|(\alpha_1-\alpha_2)-(\beta_1-\beta_2)|, (\alpha_1-\alpha_2)+(\beta_1-\beta_2)]$. \\
-- By looking at the convergence properties of the integral (\ref{Jn}) and of its derivatives, one concludes that for $n\ge 3$, $\CJ$ is a piecewise polynomial function of $\gamma$  of differentiability class {$C^{n-3}$.\\
-- In particular, for $n=3$, a simple, piece-wise linear expression may be written for $\CJ(\alpha,\beta;\gamma)$ that shows explicitly where the lines of non-differentiability lie, see \cite{Z1} and Fig.~\ref{fig-SU3} for an example. 
The resulting formulae reproduce very well the histograms obtained by numerical simulations. 
\begin{figure}[!tbp]
  \centering
  \includegraphics[width=10pc]{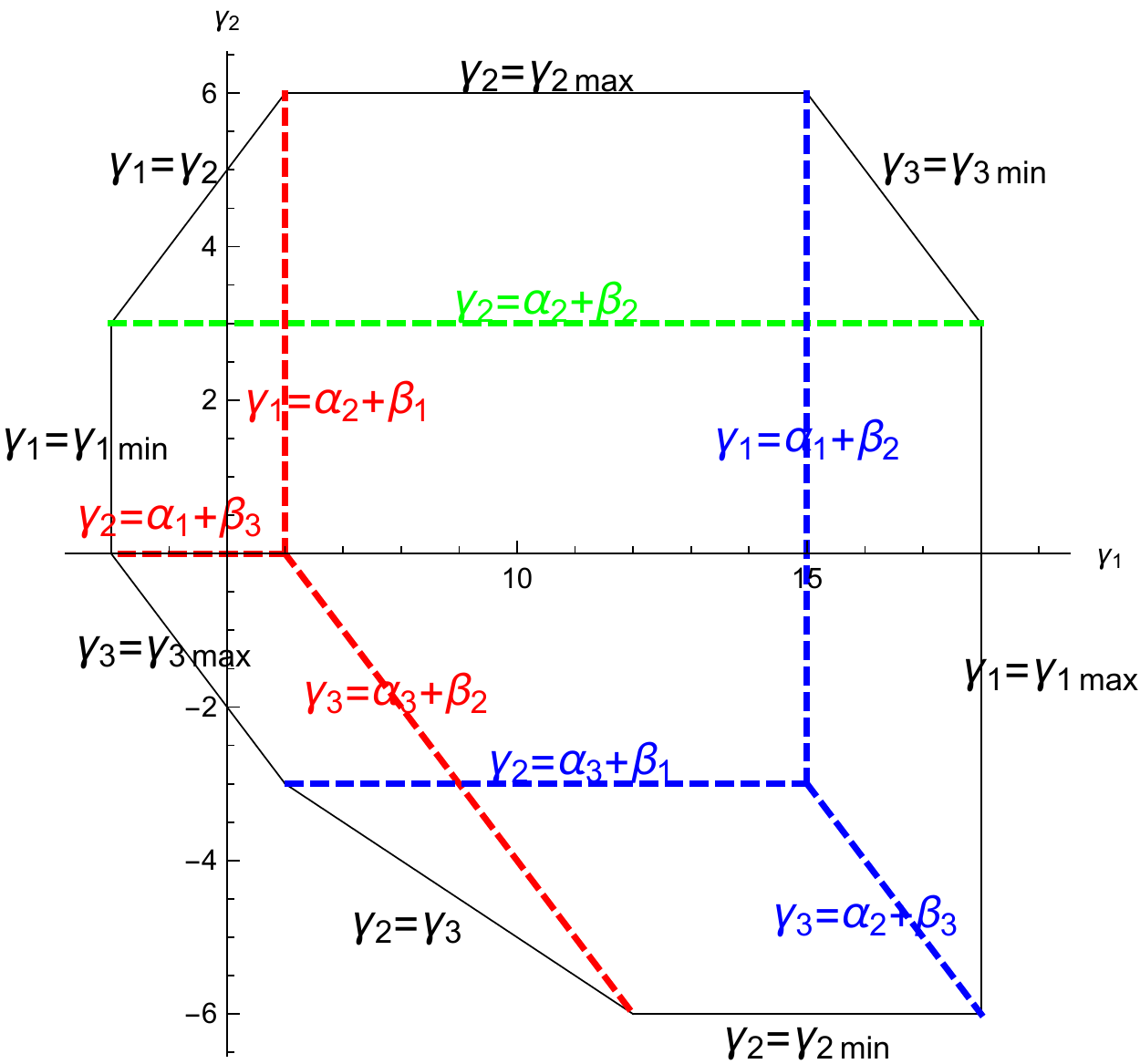}\qquad
  \includegraphics[width=12pc]{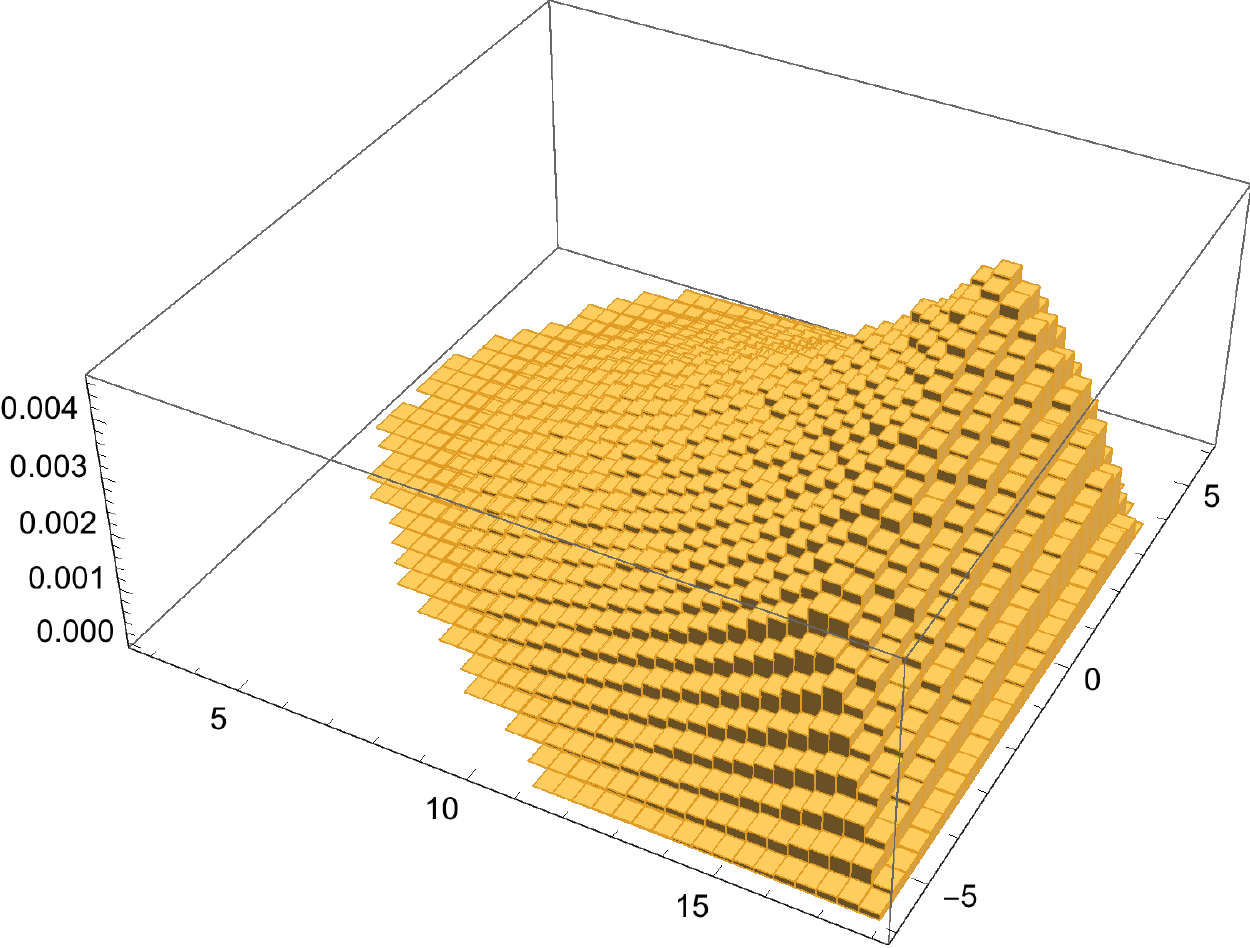}\qquad\includegraphics[width=12pc]{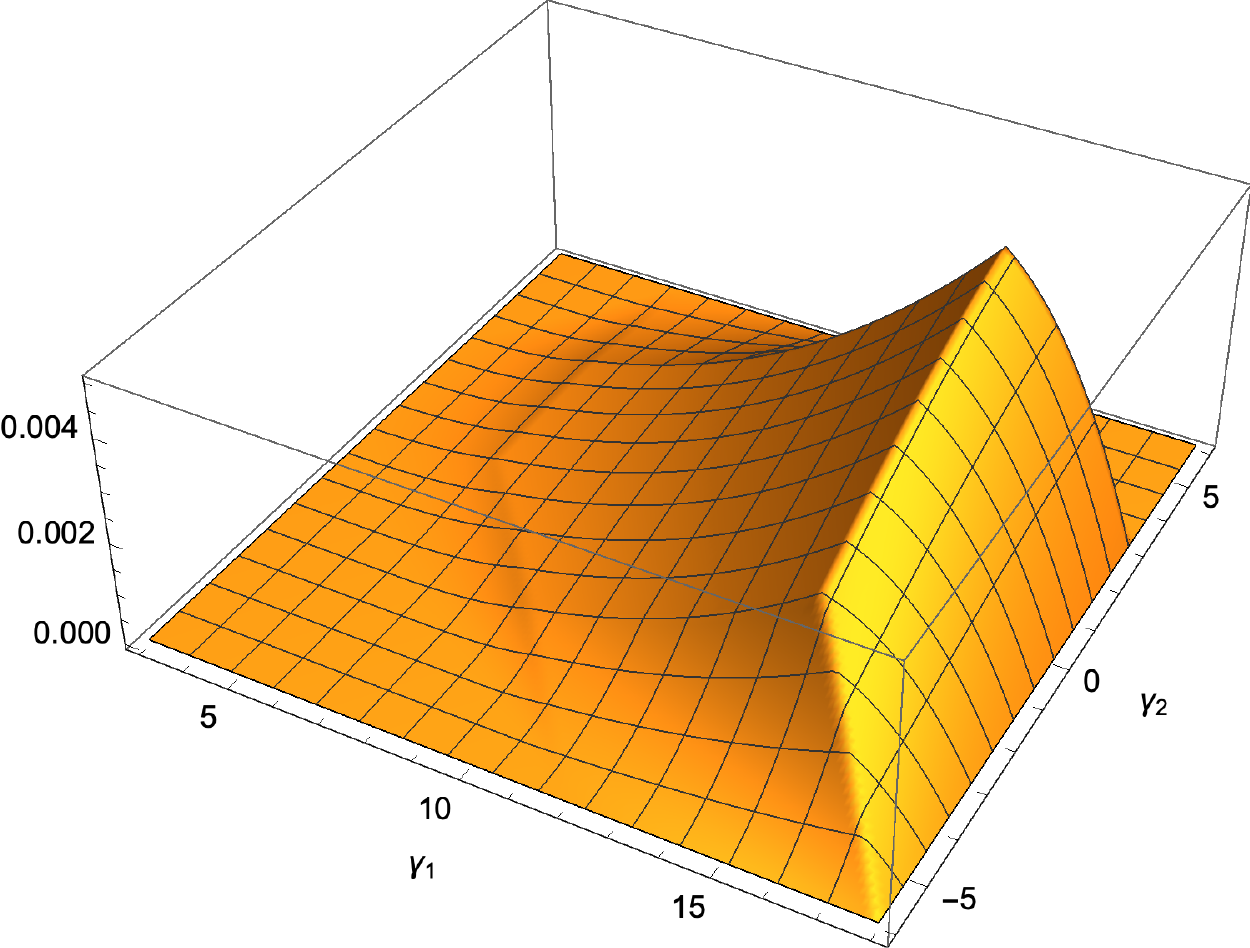}
  \caption{\small SU(3): (left) Horn's polygon with the lines of non-differentiability, drawn here for $\alpha=(11,-1,-10),\ \beta=(7,4,-11)$;
  (middle) histogram of $10^6$ samples of $\gamma$; (right) the PDF of (\ref{p-Jn}) for the same $\alpha$ and $\beta$. 
  }
  \label{fig-SU3}
  \end{figure}

\subsection{Extension to other coadjoint  representations or to quaternionic self-dual matrices}
Making use of the expressions given in sect. \ref{sect-orb-int} for the orbital integrals, and using the same method as in the previous subsection
of reduction to generalized Dirichlet integrals, the PDF may also be computed analytically for coadjoint orbits of other low-rank 
Lie algebras, or for the (self-adjoint case of)  quaternionic self-dual matrices.  One finds a PDF that is a function of differentiability class $C^{2r-3}$ for the coadjoint orbits of the  $B_r$ algebra, and $C^{2(n-2)}$ for  quaternionic self-dual $n$-by-$n$ matrices.  A sample of comparisons with numerical data
is displayed on Fig.~\ref{sample-of-res}.

\begin{figure}[!btp]
  \centering
     \includegraphics[width=0.3\textwidth]{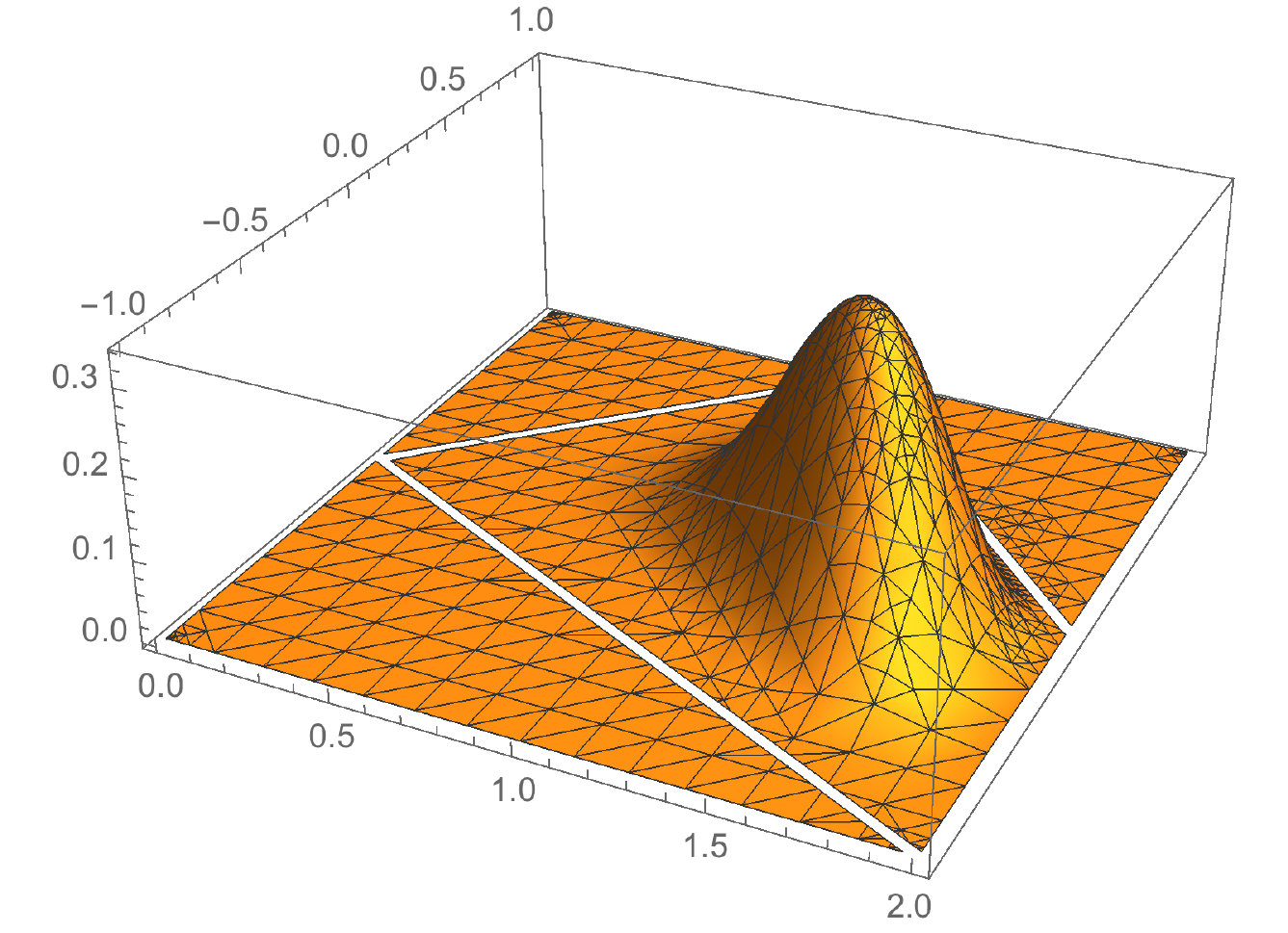}\qquad
      \includegraphics[width=0.35\textwidth]{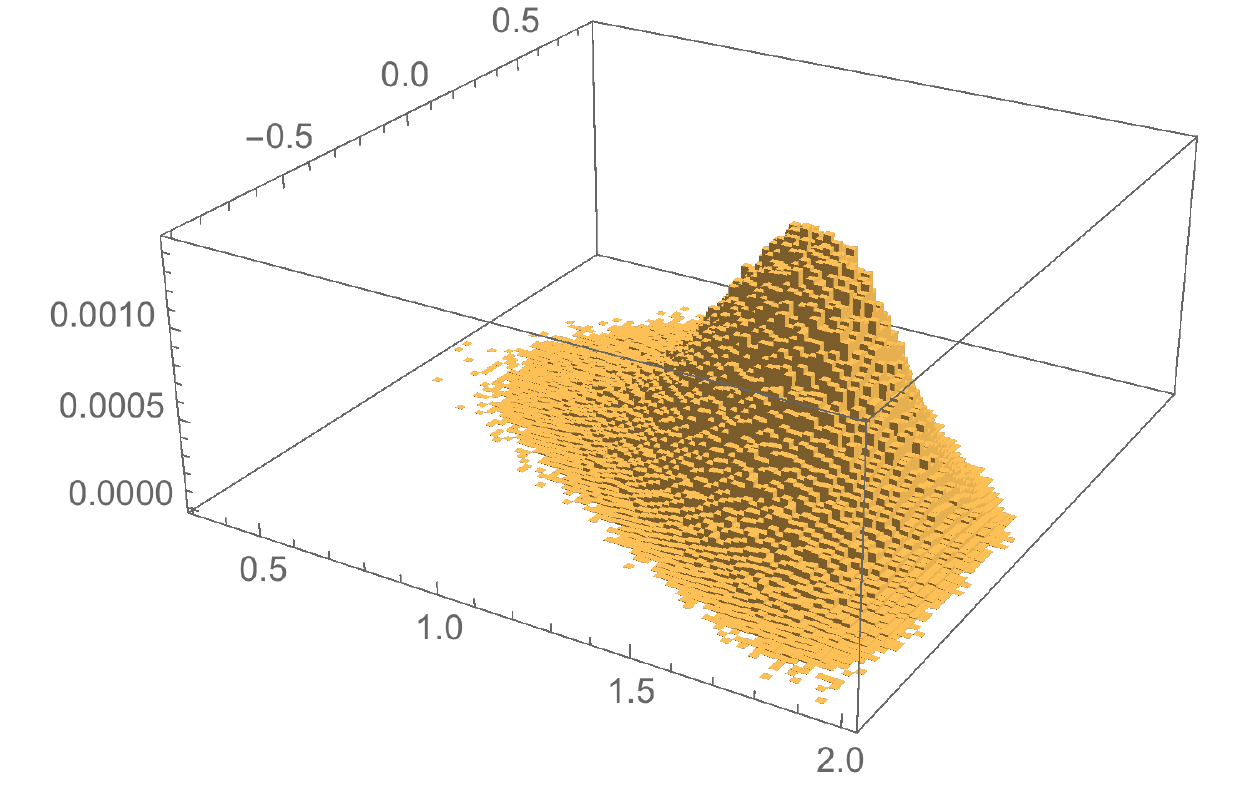}\\[10pt]
\raisebox{-1ex}{\includegraphics[width=0.3\textwidth]{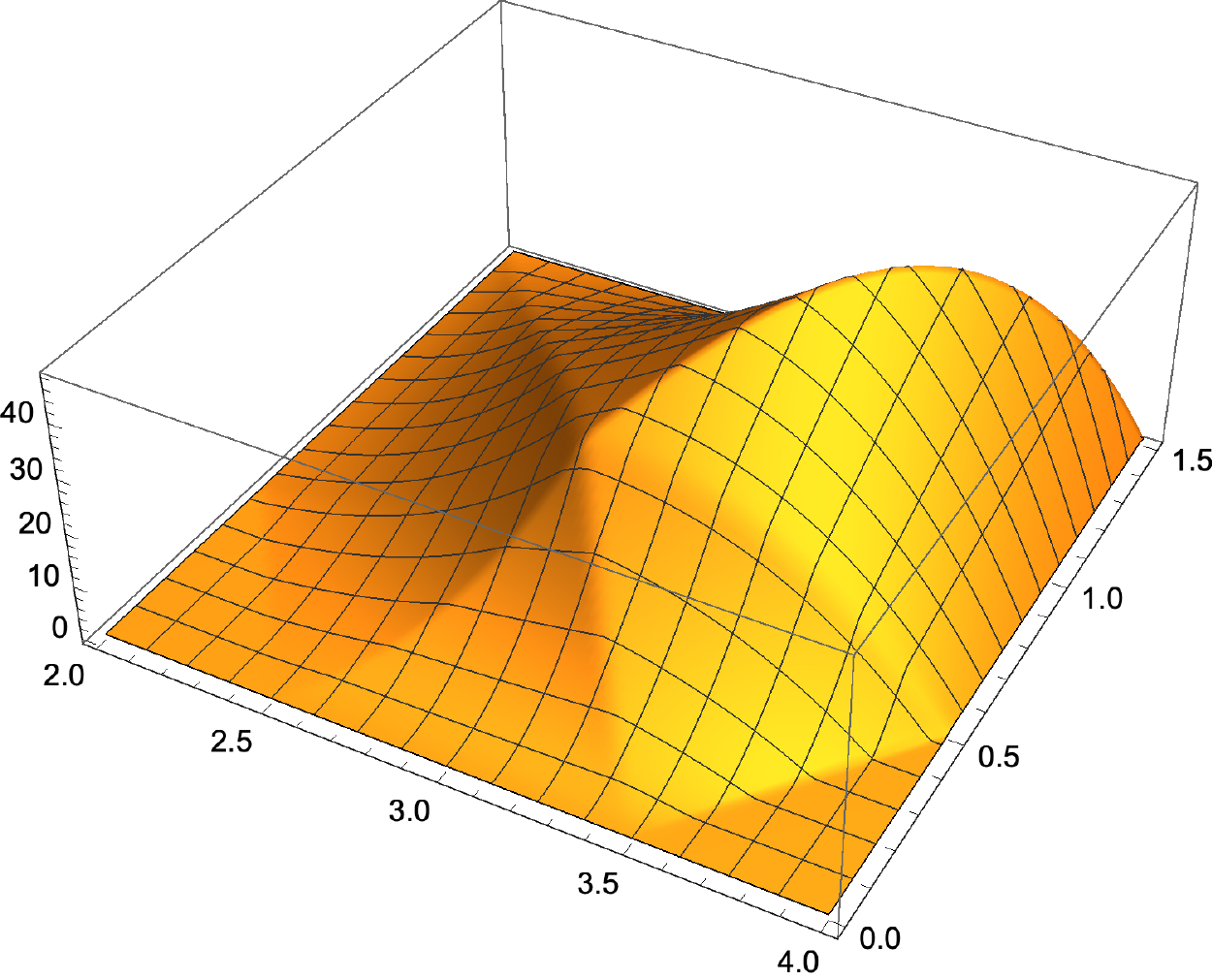}}\qquad  \includegraphics[width=0.3\textwidth]{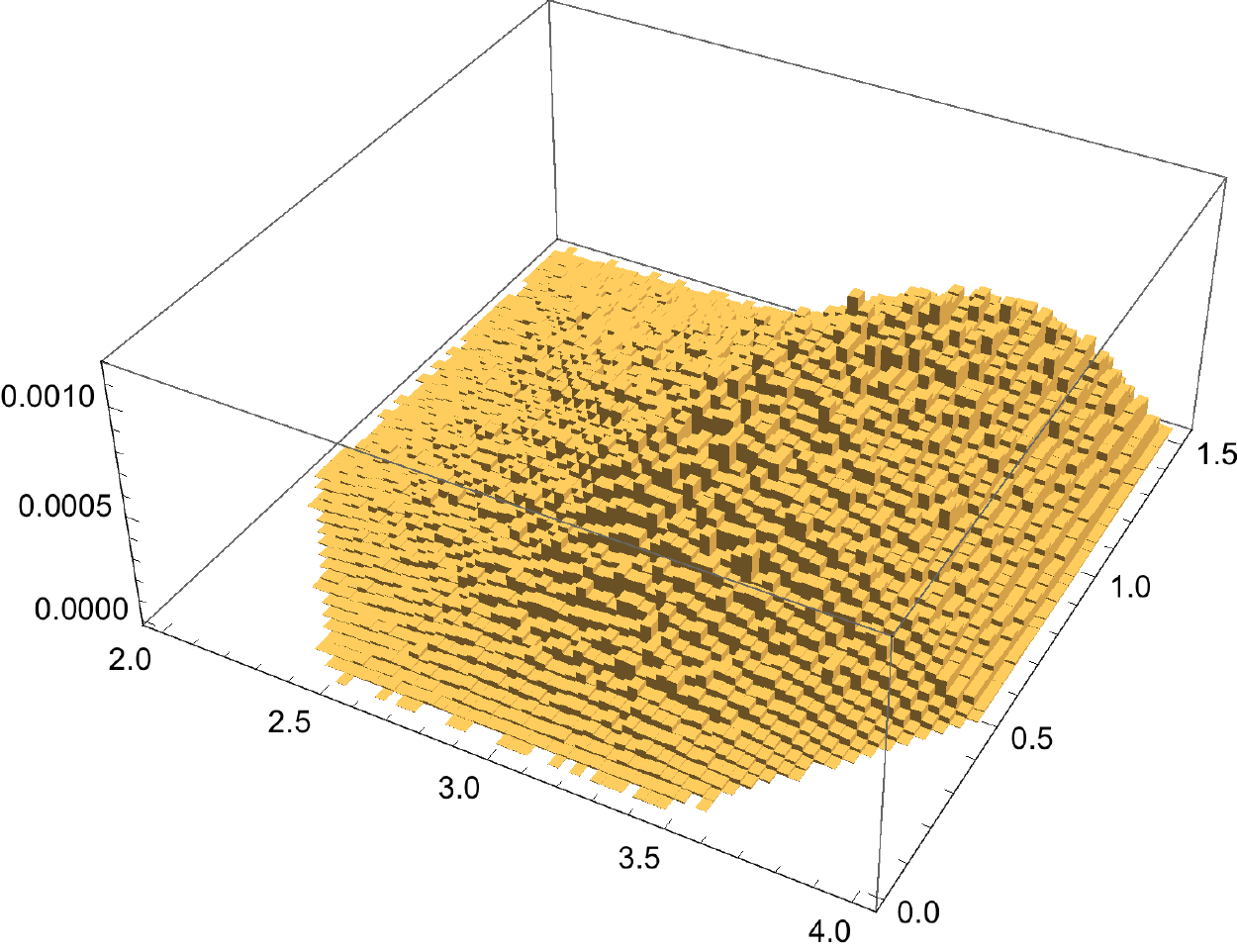}
    \caption{\label{PDFQuaternionic} \small Comparing results of analytical   calculations with ``experimental"   
    histograms. Quaternionic self-dual $3\times 3$ matrices with $\alpha=\beta=(1,0,-1)$;
    Skew-symmetric $5\times 5$  matrices with $\alpha=(1.01,1),\ \beta=(3,.5)$. }\label{sample-of-res}
\end{figure}


\def\I{I}

\def\E{\Bbb{E}}

\subsection{$\SO(2)$ and $\SO(3)$ orbits of real symmetric matrices}
The case of $\SO(n)$ acting on real symmetric matrices is both more challenging, since no manageable expression exists 
for the orbital integrals, and more intriguing, in view of the strong singularities apparent from numerical data, see Fig.~\ref{comparison}-left.

In the case of $2\times 2$ real symmetric matrices and the action of $\SO(2)$, it is a classroom exercise to work out
the PDF as a function of the differences $\alpha_{12}:=\alpha_1-\alpha_2$, $\beta_{12}:=\beta_1-\beta_2$, $\gamma_{12}:=\gamma_1-\gamma_2$.
One finds 
\be\label{PDFSO2} p(\alpha,\beta|\gamma)=\begin{cases}\frac{2}{\pi} \frac{\gamma_{12}}{\sqrt{(\gamma_{12M}^2-\gamma_{12}^2)(\gamma_{12}^2-\gamma_{12m}^2) }}& \mathrm{for}\ \gamma_{12}\in [\gamma_{12m},\gamma_{12M}]:=[|\alpha_{12}-\beta_{12}|,\alpha_{12}+\beta_{12}], \\
0 & \mathrm{otherwise,}\end{cases}\,
\ee
which exhibits an inverse--square-root singularity at the end points of its support (only the upper one if $\alpha_{12}=\beta_{12}$).
 
In ref. \cite{CZ2}, the case of orbits of real symmetric matrices under the action of $\SO(3)$ has been treated in detail.
With no loss of generality, one may assume that the matrices $A$ and $B$ are traceless. 
Then, to circumvent the lack of an expression for the orbital integral, it was found useful to trade the three eigenvalues
$\gamma_i$ (of vanishing sum) for their symmetric functions $p=\gamma_1\gamma_2+\gamma_2\gamma_3+\gamma_3 \gamma_1$,
$q=-\gamma_1\gamma_2\gamma_3$. Write the characteristic polynomial of
$C=\diag(\alpha)+\CR\, \diag(\beta)\,\CR^T$, with $\CR\in \SO(3)$, as 
$$ \det(z\, \Bbb{I}_3 - C)= z^3 + P(\CR) z +Q(\CR)\,. $$
 For given $\alpha$ and $\beta$, and $\CR$ regarded as a random variable uniformly distributed in SO(3) (in the sense of 
  Haar measure), $P(\CR)$ and $Q(\CR)$ are also random variables, whose ``PDF'' may be written formally as 
$$ \rho(p,q) =\Bbb{E}\big(\delta(P-p) \delta(Q-q)\big) =\int D\CR\,\delta(P(\CR)-p) \,\delta(Q(\CR)-q)\,. $$
Parametrize  $\CR$ in terms of Euler angles, 
 $ \CR=\CR_z(\phi) \CR_y(\theta) \CR_z(\psi)$   
with 
$ 0\le {\phi}\le 2\pi,\ 0\le {\theta} \le \pi,\  0\le {\psi}\le 2\pi $, 
and the normalized Haar measure equal to $D\CR= \inv{8\pi^2} \sin \theta\, d\theta\, d \phi \, d \psi \,.$ 
Then  $P_p:= P(\CR)-p $ and $Q_q:=Q(\CR) -q$ are degree 2 polynomials in $c=\cos \theta$, so that
 $$ \rho(p,q)=\inv{2\pi^2} \int_0^{\pi} d\phi  \int_0^{\pi} d\psi \int_{-1}^1 dc \,  \delta(P_p)   \delta(Q_q)\,,$$
while the  PDF for the independent variables $\gamma_1,\gamma_2$ is
\be\label{PDFp} \p(\gamma_1,\gamma_2)= |\Delta(\gamma)| \,\rho(p,q)\,.\ee
A curious (and apparently original) identity on delta functions of polynomials then comes to the rescue:
\be \label{eqn:delta-identity} \int dc\, \delta(P_p(c))\, \delta(Q_q(c))  = |J|\, \delta(R) \ee 
where $R$ is the {\it resultant} of $P_p(c)$ and $Q_q(c)$,
       and $J$ some Jacobian. For the conditions of applicability and proof of (\ref{eqn:delta-identity}), see \cite{BZ}.
       
    In the present case, the resultant $R$ of $P_p(c)$ and $Q_q(c)$ is a fairly big degree {4} polynomial  of 
 {$u=\cos^2 \phi$} and {$z=\cos^2 \psi$}  and $J$ is a degree 1 polynomial in $z$ and $u$, thus 
       {\bea\nonumber \rho(p,q)       &=& \inv{2\pi^2} \int_0^1 dz \int_0^1 du\, |J(u,z)|\, \delta(R)\\
       \label{rhoformula} &=& \inv{2\pi^2} \int_0^1 dz   \sum_{i\atop\mathrm{roots}\ u_i(z) \in [0,1]}   \frac{|J(u_i,z)|}{|R'_u(u_i,z)|}\,. \eea}

The calculation has been carried out in detail  in \cite{CZ2} in the particular case of $\alpha=\beta=(1,0,-1)$. Then
\be\label{rho-expl}
 \rho(p,q)= \frac{2}{\pi^2}\int_{0}^1 dz  \sum_{\mathrm{roots}\ u_i \ {\rm of}\ R\atop 0\le u_i \le 1}  \frac{(2+u+ z)}{|R'_u| }\Big|_{u=u_i}\,.\ee
 Even in that particular case, which has special symmetries, e.g. of $R$ under $z\leftrightarrow u$ and of 
 $\rho$ under $q\leftrightarrow -q$ (\ie\ $\gamma_1\leftrightarrow -\gamma_3,\ \gamma_2\to -\gamma_2$), 
the calculation is complicated by the intricate pattern of roots $u_i(z)\in [0,1]$ of 
$R(u,z)$ within the integration interval $z\in[0,1]$,
and by the  task of determining which of the possible zeros of $R'_u(u_i,z)$ give rise to a divergent integral. At the end
of the day, the result reproduces very well the numerical histogram in the $p,q$ or in the $\gamma_1,\gamma_2$ variables, see 
Fig.~\ref{anal-histo-SO3}. 

\begin{figure}\begin{center} 
\includegraphics[width=16pc]{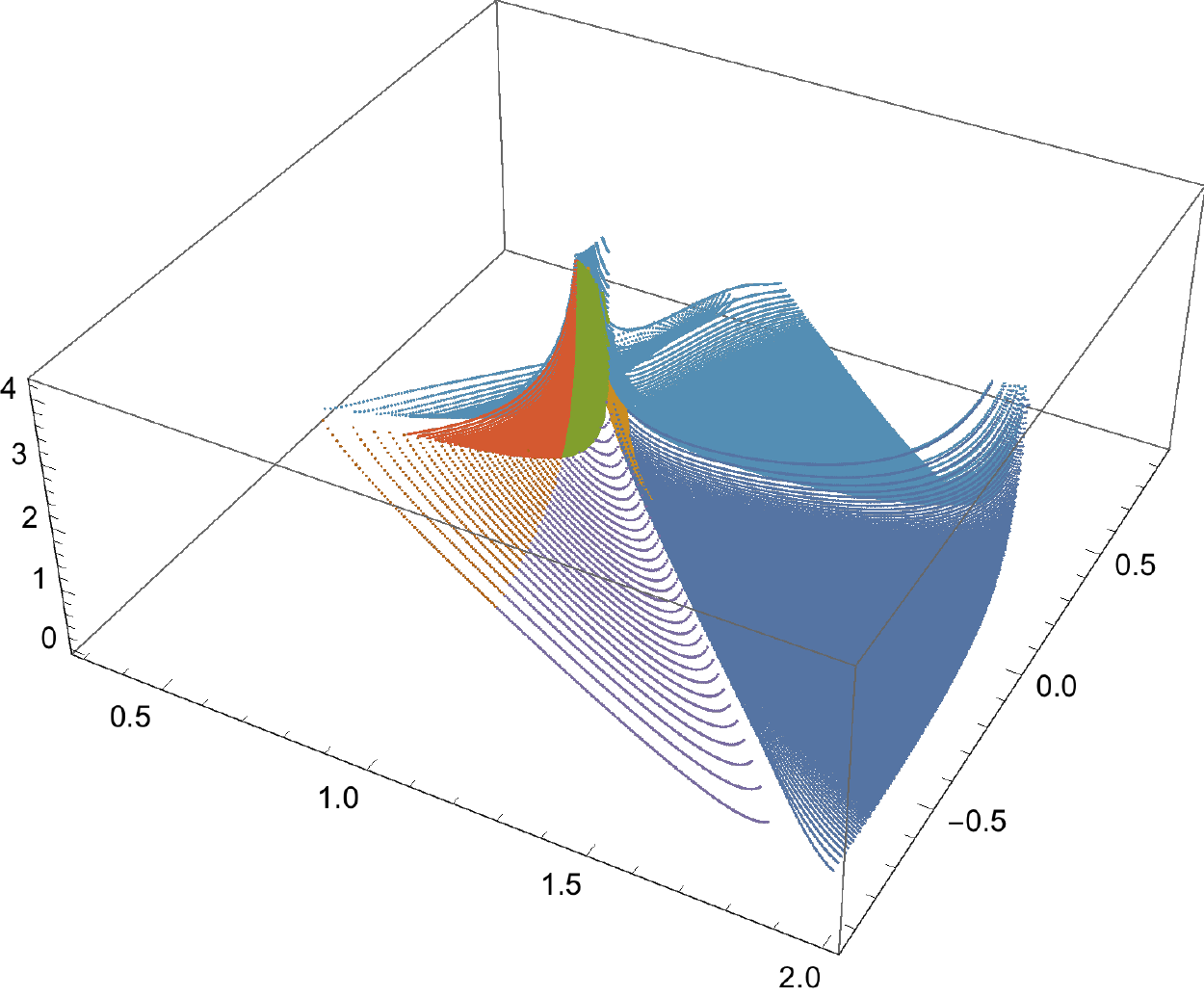}
\includegraphics[width=17pc]{histosymmJzJz1000000.pdf} 
\caption{\small Comparing the computation of $\p(\gamma_1,\gamma_2)$ for $\alpha=\beta=(1,0,-1)$ with the histogram of Fig.~\ref{comparison}}
\label{anal-histo-SO3}\end{center}\end{figure}

The main merit of the expression (\ref{rho-expl}) is  to allow a detailed discussion of the various singularities of the integral. 
Divergences of  $\rho$ can arise in two ways: 

-- From the vanishing of  $R'(u_i(z),z)$ 
at some $z_s$ by coalescence of two roots
 $u_i =u_j$ of $R$,   giving rise to  a non-integrable singularity  of $1/|R '_u|$ at $z_s$, or
 
 -- From the overall vanishing of $R'_u$ in some limit.
 
 One finds a logarithmic divergence 
of  the PDF $\p(\gamma_1,\gamma_2)$ (see eq. (\ref{PDFp}))  as $\gamma$ approaches the blue, red and magenta lines in Fig. ~\ref{Fig-div}, and 
  inverse--square-root divergences at the two special points $(\gamma_1,\gamma_2)=(1,0)$ and $(2,0)$.
  \begin{figure}\begin{center}\includegraphics[width=14pc]{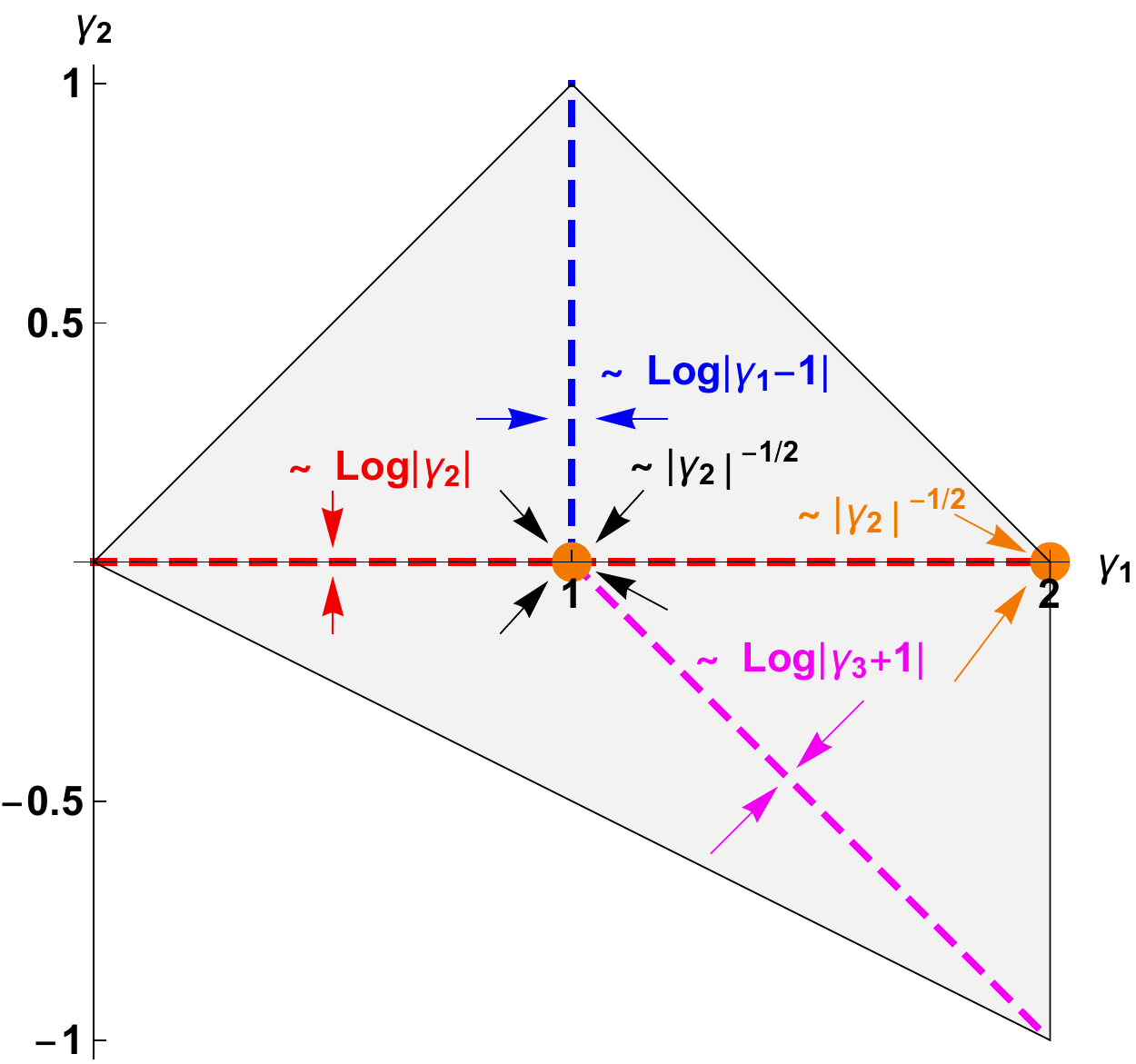}\caption{\label{Fig-div}
   The  various divergences of $p(\gamma_1,\gamma_2)$}\end{center}\end{figure}
Note that because of the vanishing of the Vandermonde determinant, $\p(\gamma_1,\gamma_2)$ may have a smaller locus of singularity than $\rho$, see \cite{CZ2}. 

\bigskip
{\bf Discussion.} 
Though it is gratifying to have reproduced the pattern and the nature of singularities of $\CJ$ (in a particular case, but this will generalize
to arbitrary $\alpha,\beta$, at the price of heavier algebra), 
this computation sheds limited light on the geometric origin of these singularities and on what should be 
expected for higher $n$. In \cite{CMSZ} it  has been argued that the  singularities find their origin in the projection from the original
$\CO_\alpha \times \CO_\beta$ to $\CC_+$,  and that they can be understood as arising from the singularity of certain coordinates on the product of orbits. The argument, unfortunately, 
says nothing about the nature of the singularity.  For higher $n$, 
should the singularity become softer and become just a non-analyticity like in the coadjoint cases,
or should a divergence persist? Numerical experiments indicate a sharp pattern of the PDF for $n=4$, but its 
precise nature remains elusive. There is clearly room for further progress.

\def\N{N}
\section{Horn's problem, Representation theory, and Combinatorics}
In this section, we shall discuss the relationship between Horn's problem and a basic problem in representation theory:
the decomposition of tensor products of representations of a compact Lie group $G$. The fact that 
multiplicities in such a decomposition
admit, for large representations, a {\it semi-classical} description has been known for a long time, see \cite{He82}. 
More specifically, we consider here the so-called ``Littlewood--Richardson (LR) multiplicities"  $ \N_{\lambda\,\mu}^\nu$,
 which appear in the decomposition of the tensor product of two irreps of highest weights $\lambda$ and $\mu$,
\be \label{tensormult}V_\lambda\otimes V_\mu =\bigoplus_\nu \N_{\lambda\,\mu}^\nu V_\nu\,.\ee
Here we will assume that $G$ is simply connected, so that we can identify representations of $G$ and $\gog$.  The bottom line is that Horn's problem appears as a {semi-classical} approximation of the LR multiplicities, as we will now explain. 

\subsection{Relation $\CJ$--LR}
The reader may have noticed the similarity between the general form of $\CJ$ in (\ref{Jn}), as an integral of the product of three $\CH$'s, and the classical expression of LR multiplicities in terms of characters of the group
\be\label{char-form} N_{\lambda\mu}^\nu= \int_G dg\, \chi_\lambda(g)\chi_\mu(g)\chi_\nu(g)^* =
\int_{\Bbb{T}} dT\, \chi_\lambda(T)\chi_\mu(T)\chi_\nu(T)^*\,,\ee
where in the second expression, the integration has been reduced to a Cartan torus $\Bbb{T}$. 
The parallel is made much tighter if one realizes that the H-C orbital integral $\CH_\gog(\alpha,\ii x)$ is  proportional to the 
character $\chi_{\lambda}(T)$, when evaluated 
 at $\alpha=\lambda+\rho$, $\rho=\oh\sum_{\Ga>0} \Ga$, and $T=e^{\ii x}\in\Bbb{T}$:
\be\label{CJ-chi}  \frac{\chi_\lambda(T)}{\dim V_\lambda}=\frac{\Delta_\gog(\ii x)}{\hat\Delta_\gog(e^{\ii x})} \CH(\lambda+\rho, \ii x) \ee
where  $ \Delta_\gog(x)$ has been introduced above in (\ref{Deltag}) and $\hat\Delta_\gog(e^{\ii x}):=  \prod_{\Ga >0} 
\Big(e^{\frac{\ii}{2}\langle \Ga,x\rangle}- e^{-\frac{\ii}{2}\langle \Ga,x\rangle  }\Big)$ is the famous ``Weyl denominator."
This remarkable formula reflects a  deep correspondence between (``classical") coadjoint orbits 
 and (``quantum") irreps of $G$: this is the object of Kirillov's orbit theory~\cite{Kir}.
 
 \def\got{\mathfrak{t}}
 \def\x{x}\def\c{r}
 
 Using that relation, one may rewrite $\CJ$ as expressed in (\ref{Jn}), evaluated at a triple of h.w. $\lambda,\mu,\nu$ shifted by $\rho$,
 as an integral of characters. We assume that the  triple $(\lambda,\mu,\nu)$ is such that $\lambda+\mu-\nu\in Q$, where $Q$ is the root lattice, 
since otherwise the LR multiplicity $N_{\lambda \mu}^\nu$ vanishes, as is well known. 
 
 The main difference between the integrals appearing in (\ref{Jn}) and 
 (\ref{char-form}) is that the former runs over the whole Cartan subalgebra $\got\simeq \R^r$, while the latter is over the (compact) Cartan torus
$\Bbb{T}$. But since $\Bbb{T} \corr{\,\simeq\,} \got/(2\pi P^\vee)$, where $P^\vee$ is the {\it coweight} lattice,
 generated by the coweights\footnote{Here the fundamental weights satisfy 
 $2\langle \omega_i,\alpha_j\rangle/\langle \alpha_j,\alpha_j\rangle=\delta_{ij}$, while the coweights are normalized so as to satisfy
$ \langle \omega^\vee_i,\alpha_j\rangle=\delta_{ij}$.}  
$\omega^\vee_i$, $i=1,\cdots, r$, 
 we may write 
{\bea\nonumber {\CJ(\lambda+\rho,\mu+\rho;\nu+\rho)} &=& \dim V_\lambda \dim V_\mu \dim V_\nu 
\int_{\mathfrak{t}\simeq \R^r} d^r \x\, |\Delta_\gog(\x)|^2\,  \CH(\lambda+\rho, \ii \x)\CH(\mu+\rho, \ii \x)(\CH(\nu+\rho, \ii \x))^*\\
\nonumber          &=& \int_{\mathfrak{t}} {\normalcolor{d^r\x\,  |\hDelta_\gog(e^{\ii \x})|^2}}\, \frac{\hDelta_\gog(e^{\ii \x})}{\Delta_\gog({\ii} \x)} \, \chi_\lambda(e^{\ii \x}) \chi_\mu(e^{\ii \x})       (\chi_\nu(e^{\ii \x}) )^*\\
\nonumber 
 &=& \int_\mathfrak{t} \,\,  \underbrace{ d^r\x\,  |\hDelta_\gog(e^{\ii \x})|^2}  \,\, \sum_{\delta\in 2\pi P^\vee}   \frac{\hDelta_\gog(e^{\ii (\x+\delta)})}{\Delta_\gog({\ii}(\x+\delta))}\, \chi_\lambda(e^{\ii { (\x+\delta)}}) \chi_\mu(e^{\ii {(\x+\delta)}})  (\chi_\nu(e^{\ii {(\x+\delta)}}) )^*\\
  \label{lasteq}&=& \int_\T   \qquad    dT \quad {\normalcolor{\left( \sum_{\delta\in {2\pi} P^\vee}  e^{\ii \langle \rho, \delta\rangle} 
   \frac{\hDelta_\gog(\corr{e^{\ii \x}})}
   {\Delta_\gog({\ii}(\x+\delta))}  \right)}}\, \chi_\lambda(T) \chi_\mu(T)  (\chi_\nu(T) )^*.
 \eea}
 
Now, the sum over $\delta\in P^\vee$ is a generalization of the classical identity 
$\sum_{n=-\infty}^\infty \frac{(-1)^n}{u+(2\pi)n}=\inv{2\sin(u/2)}$
and in general yields~\cite{ER} 
\bea \label{ER-ident}\sum_{\delta\in {2\pi} P^\vee}  e^{\ii \langle \rho, \delta\rangle}  
  \frac{\hDelta_\gog(\corr{e^{\ii \x}})}{\Delta_\gog({\ii}(\x+\delta))}  &=&  
{\normalcolor\sum_{\kappa\in K}   \c_\kappa \chi_\kappa(T), } 
 \eea
where the sum on the right runs over a {\it finite, $(\lambda,\mu,\nu)$-independent} set of weights $K$
\corr{described in \cite{CZ1,CMSZ}. The coefficients $\c_\kappa$  are non-negative rational numbers satisfying}
\be\label{relation}\sum_\kappa \c_\kappa \dim V_\kappa=1\,.\ee
 For each such term, the integral over $\Bbb{T}$ thus yields the multiplicity 
$\lambda\otimes\mu\otimes\kappa\to \nu$. 
There is a similar identity involving $\CJ$ at unshifted weights $(\lambda,\mu,\nu)$~\cite{CMSZ}. 
We conclude that for a triple $(\lambda,\mu,\nu)$ such that $\lambda+\mu-\nu\in Q$, we have two identities (distinct if $\rho\notin Q$):
\bea \label{Jn-LR} \CJ(\lambda+\rho,\mu+\rho;\nu+\rho) &=&\sum_{\kappa\in K, {\tau}} \c_\kappa  N_{\lambda\,\mu}^{\tau}  N_{{\tau}\,\kappa}^{{\nu}}
   =\sum_{\kappa\in K} \c_\kappa N_{\lambda\,\mu\,\kappa}^{\,\nu} \\
\label{Jn-LRp}   \CJ(\lambda,\mu;\nu) &=&\sum_{\kappa\in \hat K, {\tau}} \hat\c_\kappa  N_{(\lambda-\rho)\,(\mu-\rho)}^{{\tau}}  N_{{\tau}\,\kappa}^{{\nu-\rho}}
   =\sum_{\kappa\in \hat K} \hat\c_\kappa N_{(\lambda-\rho)\,(\mu-\rho)\, \kappa}^{\,\nu-\rho} 
\,. \eea

For example, for SU(3), the sum in the rhs of (\ref{ER-ident}) includes only the trivial representation, so that we have simply
$\CJ(\lambda+\rho,\mu+\rho;\nu+\rho)= N_{\lambda\,\mu}^{\,\nu}$. In contrast, for SU(4) and \corr{Spin}(5) respectively, 
\bea \CJ(\lambda+\rho,\mu+\rho;\nu+\rho)&=& \inv{24} 
\Big(9 N_{\lambda\,\mu}^{\,\nu}+ N_{\lambda\,\mu\,(\omega_1+\omega_3)}^{\,\nu} \Big),\qquad 
\CJ(\lambda,\mu;\nu) =\inv{6} N_{\corr{\lambda-\rho\,\mu-\rho}\,\,\omega_2}^{\,\corr{\nu-\rho}} ,\\ 
 \CJ(\lambda+\rho,\mu+\rho;\nu+\rho)&=& \inv{8} 
\Big(3 N_{\lambda\,\mu}^{\,\nu}+ N_{\lambda\,\mu\,\omega_1}^{\,\nu} \Big),\qquad \qquad
\CJ(\lambda,\mu;\nu) =\inv{4} N_{\corr{\lambda-\rho\,\mu-\rho}\,\,\omega_2}^{\,\corr{\nu-\rho}}\,.
\eea
 
\subsection{The BZ polytope, its stretching,  and the $\CJ$ function as a volume}
We now change gear and introduce combinatorial methods to determine the LR coefficients. This follows from the 
work of Berenstein and Zelevinsky~\cite{BZ}, who have shown that given a triple $(\lambda,\mu,\nu)$ such that 
$\sigma:=\lambda+\mu-\nu\in Q$,  one may construct a 
\corr{polytope $H_{\lambda\,\mu}^\nu$,}
such that the LR coefficient $N_{\lambda\,\mu}^\nu$ is given by the number of integer points in  $H_{\lambda\,\mu}^\nu$.
\corr{We consider this polytope as a subset of $\R^k$ where $k$ is the number of positive roots; its dimension is at most $d = k - \text{rank}(G)$.
Then }
$$N_{\lambda \mu}^\nu = \# ( H_{\lambda \mu}^\nu \cap \Z^k ).$$
We call $H_{\lambda\,\mu}^\nu$ the {\it BZ polytope} associated to the triple $(\lambda, \mu, \nu)$. In general it is rational but not integral (i.e. the coordinates of its vertices are rational but not always integers).  Moreover, for $\nu$ on the interior of the support of $\CJ$, the dimension $d$ coincides with the degree of homogeneity of $\CJ$.

Determining the number of integer points in a rational polytope and its dilations is a classical problem in combinatorics. If $sH_{\lambda\,\mu}^\nu$ is the dilated polytope $\{ sx \, :\, x \in H_{\lambda \mu}^\nu \}$ for some positive integer $s$, then the number of integer points in $sH_{\lambda\,\mu}^\nu$ is given by the {\it Ehrhart quasi-polynomial} 
$$P_{\lambda\,\mu}^\nu(s) =  \# ( sH_{\lambda \mu}^\nu \cap \Z^k ) = \sum_{\ell=0}^d   s^\ell a_\ell(s),$$
where ``quasi'' means that the coefficients $a_\ell(s) $ may be periodic functions of $s$. One can prove that whenever $H_{\lambda \mu}^\nu$ is an integral polytope,
$P_{\lambda\,\mu}^\nu(s)$ is an honest polynomial. But there are cases where  
$P_{\lambda\,\mu}^\nu(s)$ is a polynomial even though $H_{\lambda \mu}^\nu$ is not integral; indeed the LR stretching polynomials for $\SU(n)$ are always honest polynomials even for non-integral BZ polytopes. 
In contrast, for the 
\corr{Spin}$(2m+1)$ groups (\ie for the $B_m$ algebra\corr{s}), one encounters many cases of quasi-polynomiality, though we have not yet been able to discern a criterion determining which $P_{\lambda \mu}^\nu$ are actual polynomials.

Using standard results in combinatorics, it can be shown that the 
coefficient $a_d$ of $P_{\lambda \mu}^\nu$ is a constant equal to the \corr{$d$-}volume of $H_{\lambda \mu}^\nu$.  (In fact it is the {\it relative} volume, given by the Euclidean $d$-volume times a scalar factor that can be computed, see \cite{CMSZ} for details.)
We now want to show that this volume  is given by the function $\CJ$. Writing eq. (\ref{Jn-LR}) for a triple of dilated weights 
$(s\lambda, s\mu, s\nu)$, $s\gg 1$, we have 
\bea\label{large-s} \CJ(s\lambda +\rho, s\mu+\rho;s\nu+\rho) &\approx&   \CJ(s\lambda, s\mu;s\nu)= s^{d} \CJ(\lambda, \mu;\nu)\\
\nonumber  &=&\sum_{\kappa\in K} \c_\kappa N_{s\lambda\, s\mu\, \kappa}^{s\nu}
\buildrel {s\gg 1} \over {\approx} \underbrace{ \sum_{\kappa\in K} \c_\kappa \dim V_\kappa}_{=1} N_{s\lambda\,s\mu}^{s\nu} = N_{s\lambda\,s\mu}^{s\nu} \eea 
where in the first line we have made use of the continuity (for $n>2$ in $\SU(n)$) and of the homogeneity of the function $\CJ$; in the second
line, for $s$ large, all the weights of the irrep of h.w. $\kappa$ contribute additively  and with their multiplicity to 
$N_{s\lambda\, s\mu\, \kappa}^{s\nu}\approx \dim V_\kappa N_{s\lambda\,s\mu}^{s\nu} $, and we use the relation 
(\ref{relation}). 

We conclude that for $s$ large
$$  s^{d} \CJ(\lambda, \mu;\nu) \approx N_{s\lambda\,s\mu}^{s\nu} =P_{\lambda\,\mu}^{\nu}(s) \approx a_d s^d = \mathrm{Vol}(H_{\lambda\,\mu}^\nu) s^d\, ,$$
whence 
\be\label{Jn-vol}  \CJ(\lambda, \mu;\nu) =\mathrm{Vol}(H_{\lambda\,\mu}^\nu)\,. \ee
This vindicates the claim that the Horn problem is a semi-classical description of the LR multiplicity problem, and that the 
function $\CJ$ measures the volume of the BZ polytope. Returning to the relations (\ref{Jn-LR}, \ref{Jn-LRp})\, we see that they give an exact expression for this volume as a finite sum of LR multiplicities, which is more precise than the previously known asymptotic relationship.


\subsection{Pictographs}
This relationship between the two problems, Horn and LR, has been beautifully illustrated by the honeycomb/hive construction
of Knutson and Tao~\cite{KT99}. KT-honeycombs are examples of pictographs, \ie of graphical combinatorial objects that describe
the two problems.

In the LR problem, the basic idea is that there should be as many (distinct) pictographs with prescribed  ``external labels'' specifying the given three highest weights  $\lambda, \mu, \nu$ (or the three irreps) as the multiplicity itself.
In other words the number of pictographs should be equal to the dimension of the space of intertwiners $\mathrm{Hom}(V_\lambda \otimes V_\mu,  V_\nu)$.
KT-honeycombs are well suited for studying the Horn problem and the GL($n$) or the $\U(n)$ multiplicity problem; their three sides are labelled by Young diagrams, i.e. by the summands of the associated integer partitions, in other words by the Young variables specifying three given irreps. 

KT-honeycombs can also be used to describe multiplicities for $\SU(n)$, but in this case three other kinds of pictographs are often better suited. 
The pictographs that we have in mind have Dynkin labels attached to their three sides. We shall distinguish three kinds of pictographs, which look different but are essentially equivalent: the Berenstein-Zelevinsky triangles (or BZ-triangles for short), the Ocneanu blades (or O-blades) and the $\SU(n)$ isometric honeycombs. The first were introduced in \cite{BZ2}, the second in \cite{AO_Oblades}, and the last were discussed, in the framework of SU(3), by two of us in sect.~4 of \cite{CZ3}. KT-honeycombs (and hives), in relation with the solution of the Horn problem in the Hermitian case, are discussed in many places, so we shall restrict this short discussion to the $\SU(n)$ case and remind the reader how the last three kinds of pictographs are related. It will be enough to present a simple example:  let us consider, in SU(4), the tensor product $V_{21,13,5}\otimes V_{7,10,12}$ where the indices (non-negative integers) refer to the Dynkin labels of two highest weights. The decomposition of this tensor product into a sum of irreducible representations contains $537186$ terms, most of them with non-trivial multiplicity because only $7092$ are inequivalent. The representation $V_{20,11,9}$, for instance, occurs with multiplicity $367$. This means that there will be $367$ distinct BZ-triangles with the given labels, and the same number of O-blades and $\SU(n)$-honeycombs. 
Fig.~\ref{threePictographs} displays one of them, in its three avatars.

  \begin{figure}\begin{center}
  \includegraphics[width=11pc]{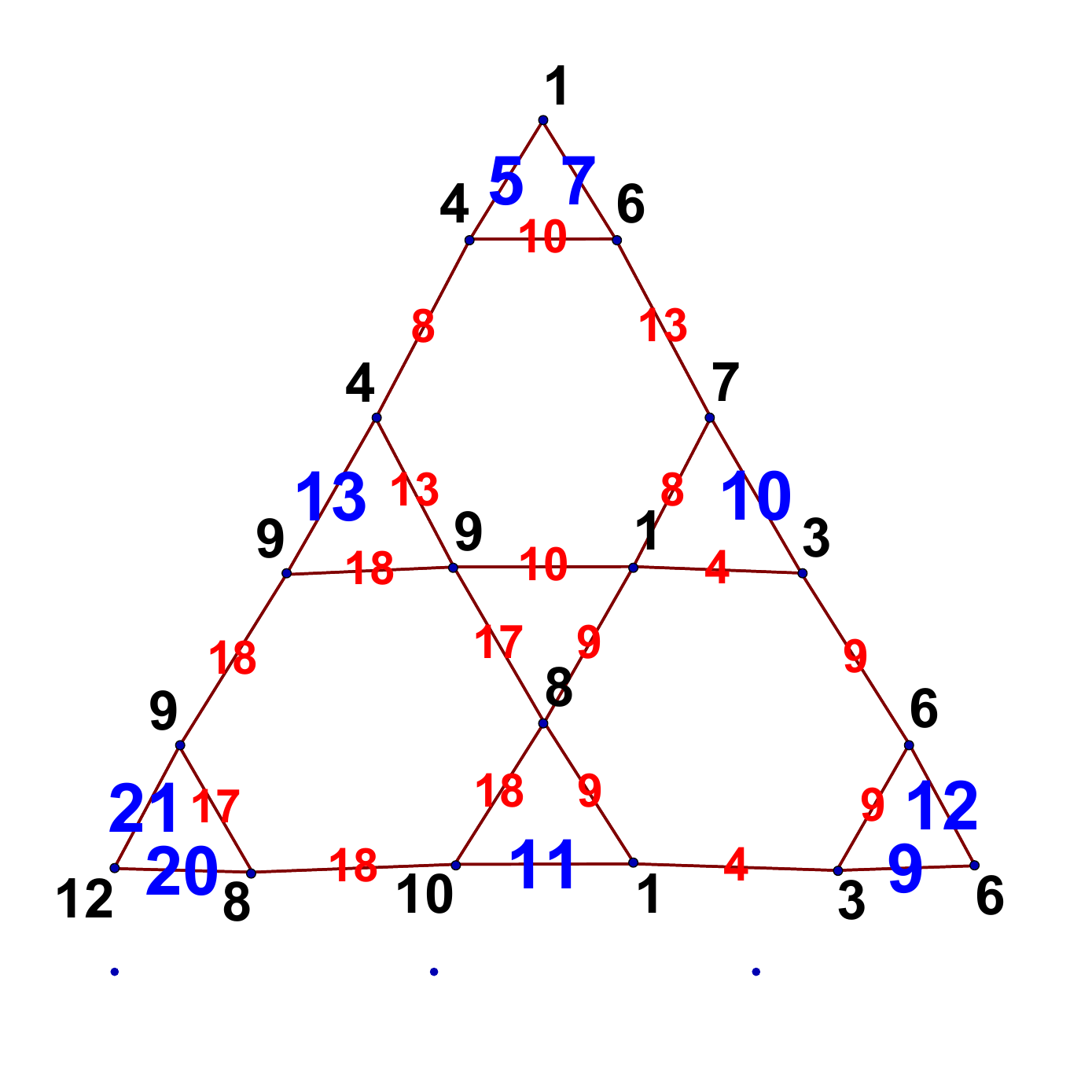}
    \includegraphics[width=12pc]{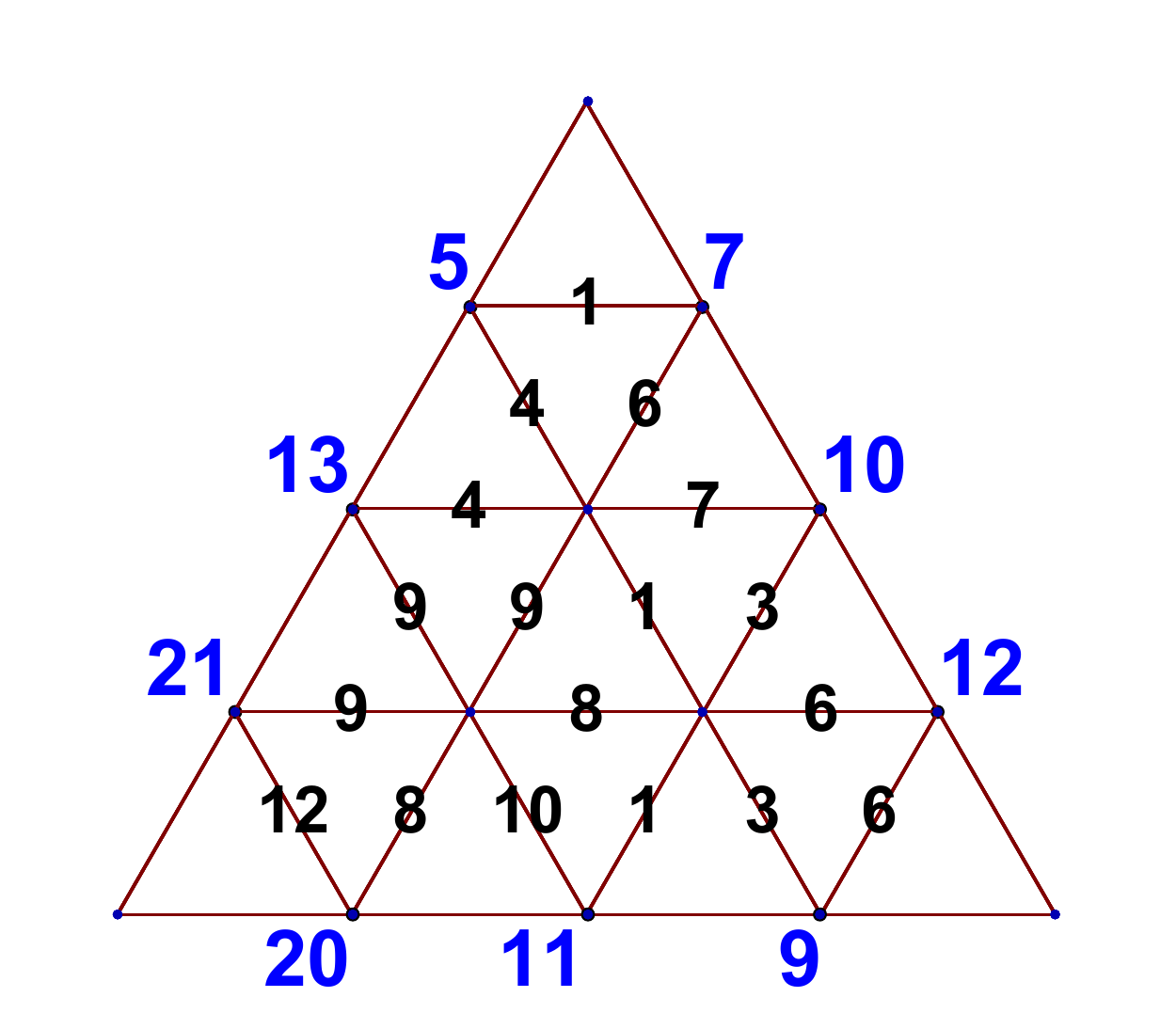}
      \includegraphics[width=12pc]{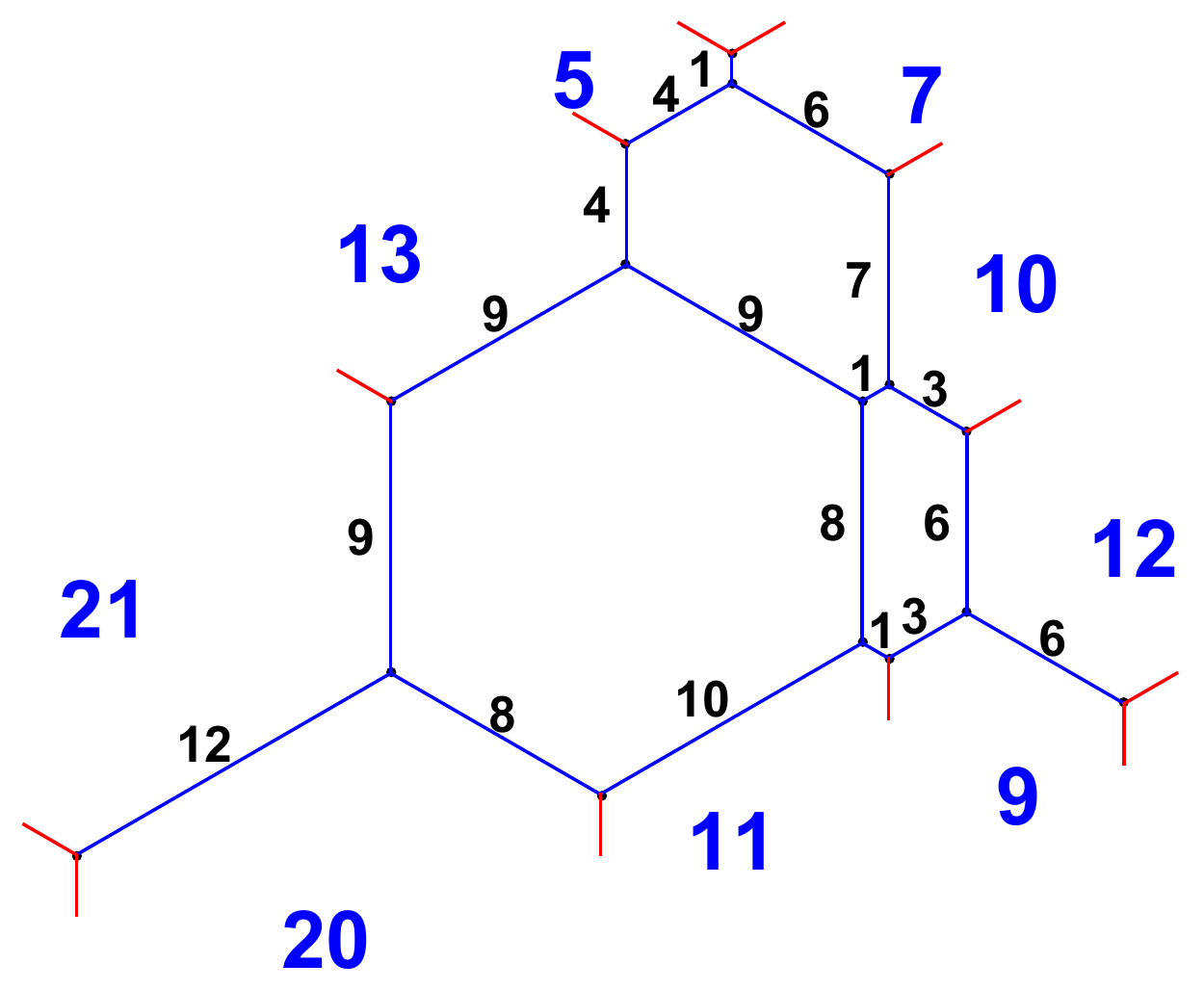}
  \caption{Three equivalent pictographs: a BZ-triangle, an O-blade, and an isometric SU(4) honeycomb.}
  \label{threePictographs}\end{center}\end{figure}
  
    \begin{figure}\begin{center}
  \includegraphics[width=18pc]{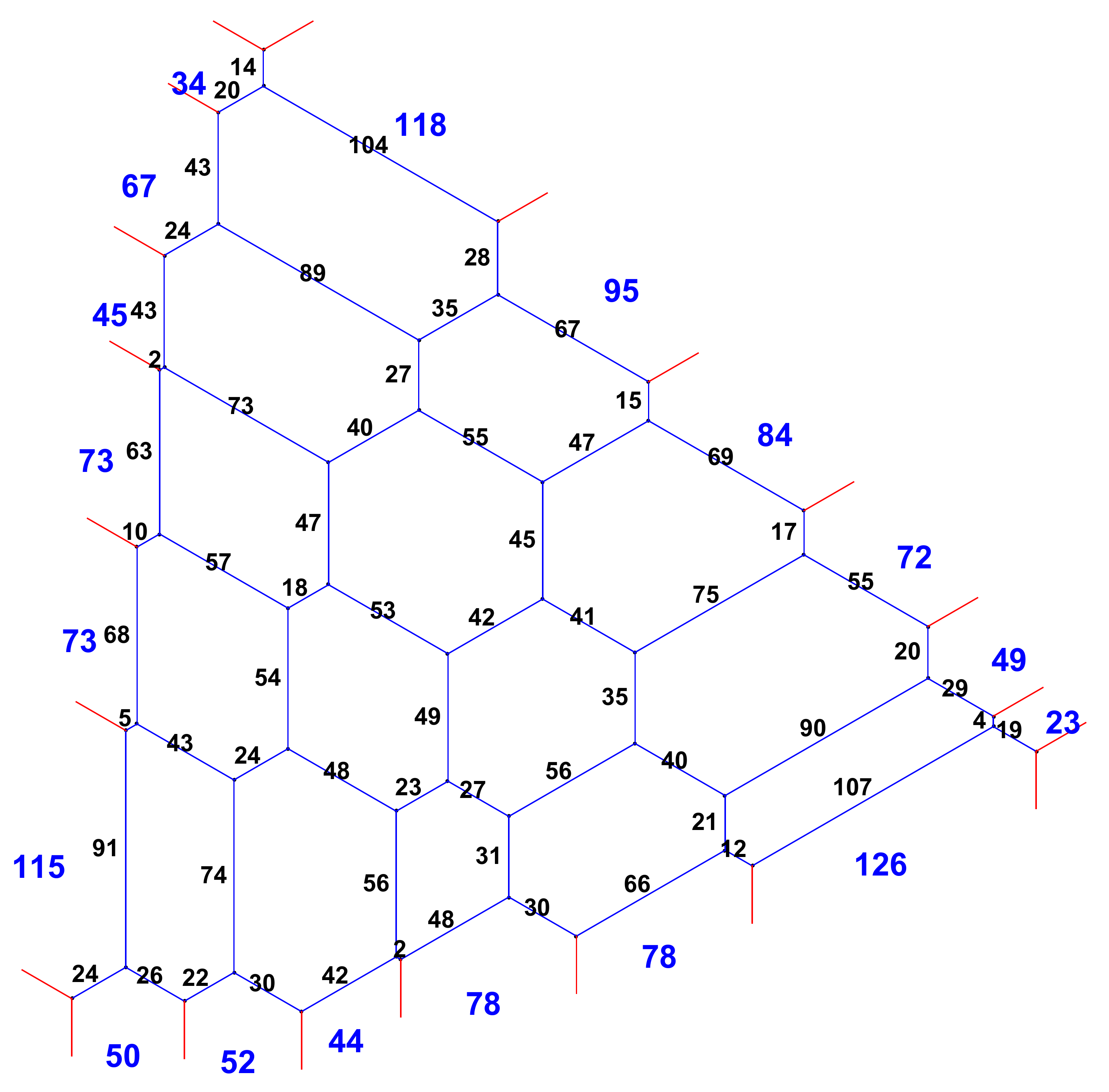}
  \caption{An isometric SU(7) honeycomb displaying one possible coupling for \\ $\{ 115,73,73,45,67,34\} \otimes \{ 118,95,84,72,49,23\}  \rightarrow \{ 50,52,44,78,78,126\}  $ .}
  \label{su7_honeycomb_example}\end{center}\end{figure}
  
In the BZ-triangle, the pattern of black integers is such that an integer carried by an edge is the sum of the integers labelling its end-points; the integers in blue are given (Dynkin labels).  Moreover, we have one additional constraint: the  red integers carried by opposite sides of hexagons are equal. In the O-blade, there is  ``conservation of the external integers'' (Dynkin labels); we did not display the red integers: they sit in the six angles surrounding the three inner vertices, and the constraint, now, is that opposite angles (defined as the sum of their corresponding edges) should be equal. In the SU(4) honeycomb the constraint is that sums of two edges relative to opposite points of each of the three hexagons are equal.

Which pictograph one prefers is a matter of taste since the geometric correspondence between the three pictures is rather obvious. In particular the honeycomb is obtained from the O-blade by a star-triangle operation, also called a  Y-$\Delta$ transform; the constraints are automatically satisfied by displaying the hexagons of the resulting honeycomb in a metric way as parallelo-hexagons (opposite sides are parallel), or, equivalently, as equiangular hexagons (each angle has a value equal to $120^\circ$), because  the length of each side is then given by the non-negative integer it carries.

 In an equiangular hexagon, the sums of two consecutive edges surrounding opposite vertices are equal (for a proof of this elementary fact, extend the six sides of the chosen hexagon and embed the latter in one of the two resulting equilateral triangles). Remember also that the black integers are non-negative, but they are allowed to equal $0$, in which case the irregular hexagons may degenerate to pentagons or to quadrilaterals.  For $\SU(n)$, there are $(n-1)(n-2)/2$ inner vertices in the O-blades (the same as the number of hexagons), and we can intuitively interpret the existence of some non-trivial multiplicity relative to a chosen triple of irreps as a kind of ``breathing'' of the (irregular) hexagons. More properties of O-blades and isometric $\SU(n)$ honeycombs, in particular their decomposition on ``fundamental pictographs,'' can be found in \cite{CZ3}.  Notice that the external sides of  the KT-honeycombs are labelled by partitions, whereas those of the $\SU(n)$ isometric honeycombs are labelled by Dynkin indices.  Moreover the numbers carried by the leaves (internal edges) are non-negative integers in the latter case, whereas they can be negative in the former (which cannot be ``isometric,'' of course).  For purposes of illustration, Fig.~\ref{su7_honeycomb_example} displays an $\SU(7)$ isometric honeycomb for a triple of highest weights with rather large entries.

Although one can construct BZ polytopes for all simple Lie groups, pictographs have been invented only for $\SU(n)$.  Finding analogs of the latter for other types of simple Lie group is a problem that has baffled the community and is still waiting for an answer.
\subsection{Relation with symmetric polynomials}

 Since characters of representations combine multiplicatively under the tensor product, the multiplicities $N_{\lambda \mu}^\nu$ can be interpreted as structure constants in the ring of symmetric functions generated by the characters of the irreps of $\gog$.  In the case of $SU(n)$ for example, these multiplicities encode the structure constants for the Schur polynomials.

In the coadjoint case, we have a notion of BZ polytopes (and, in the $\SU(n)$ subcase, we also have pictographs); we know that the volume of a BZ-polytope is measured by the value of the $\CJ$ function, a value that can be obtained by looking at the highest degree coefficient of the stretching (or LR) polynomial when multiplicities are scaled.  Moreover, we have an equality between structure constants of the algebra of symmetric polynomials in the Schur basis and the number of integer points of appropriate BZ-polytopes (or hive polytopes), for $\SU(n)$.  We obtain therefore a relation between the function $\CJ$, as defined in the (Hermitian) Horn problem, and the scaling behavior of appropriate structure constants of the ring of symmetric polynomials in the Schur basis. Clearly this kind of relation  extends to other situations, where the Lie group $\SU(n)$ is replaced by other simple Lie groups and where Schur polynomials are replaced by orthogonal Schur polynomials, symplectic Schur polynomials, etc.
 
In the self-adjoint case, however, there is no obvious notion of multiplicity and there are no BZ-polytopes. Nevertheless, we still have a $\CJ$ function stemming from the Horn problem. 
Could this $\CJ$ function be related to some kind of volume, or to some kind of asymptotic behavior  for the structure constants of an appropriate class of polynomials? The answer to the second part of the question is positive: as discussed in \cite{CZ2}, one shows that for $\theta = 1/2$ or 2, $\CJ$ is the limit of the structure constants of appropriate zonal polynomials (Jack polynomials with parameter $\theta$), see also \cite{OkounkovOlshanski}. 
The approach to asymptotics is illustrated in Fig.~\ref{zonalversusJ} which displays both the volume function $\CJ=\rho$, calculated from the integral of eq.~(\ref{rhoformula}), for some choice of its arguments, 
and a vertically scaled version of the surface approximating the corresponding\footnote{See \cite{CZ2} for more details.} zonal structure constants.

\begin{figure}[!tbp]
  \centering
   \includegraphics[width=10pc]{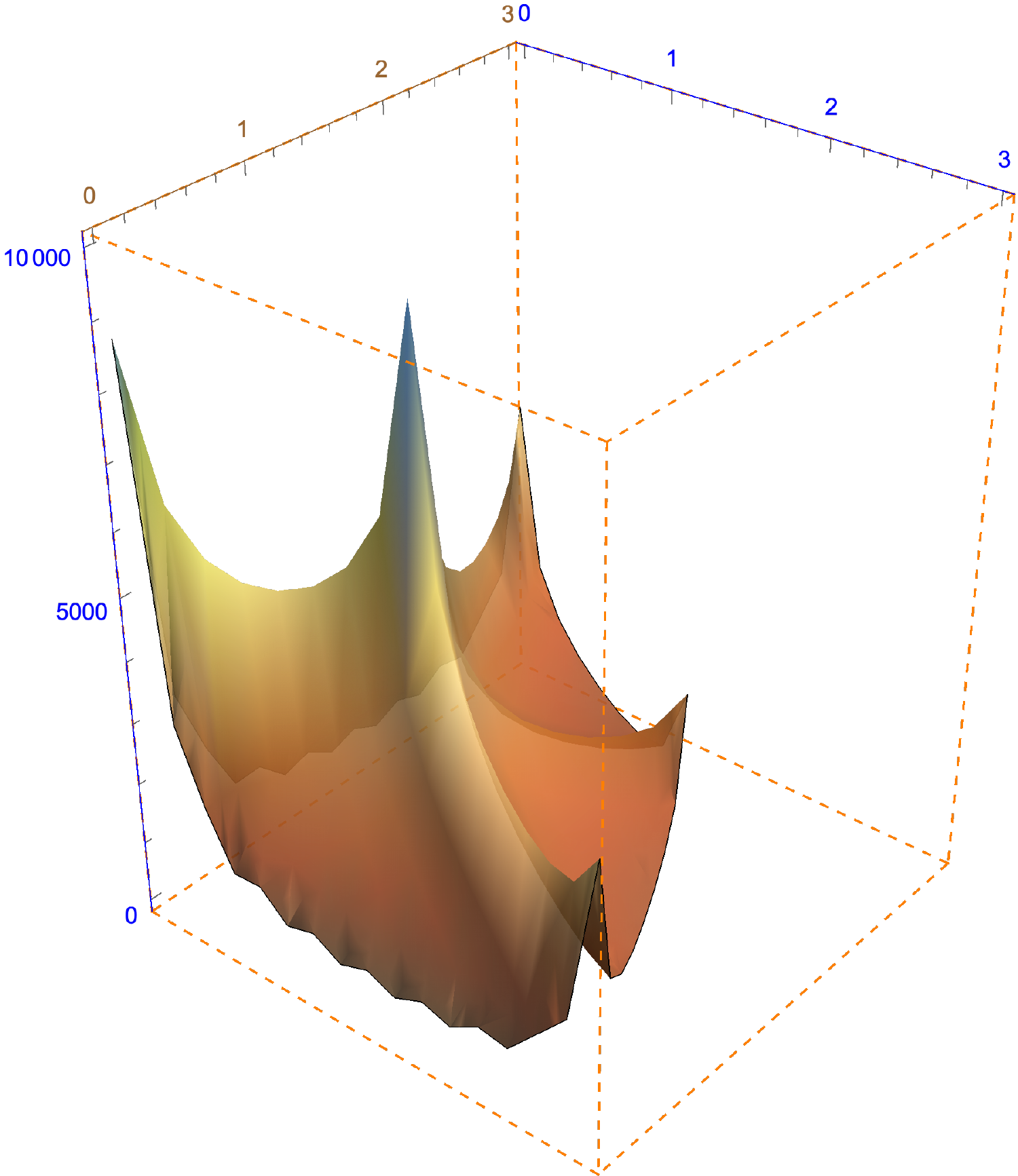}
    \includegraphics[width=10pc]{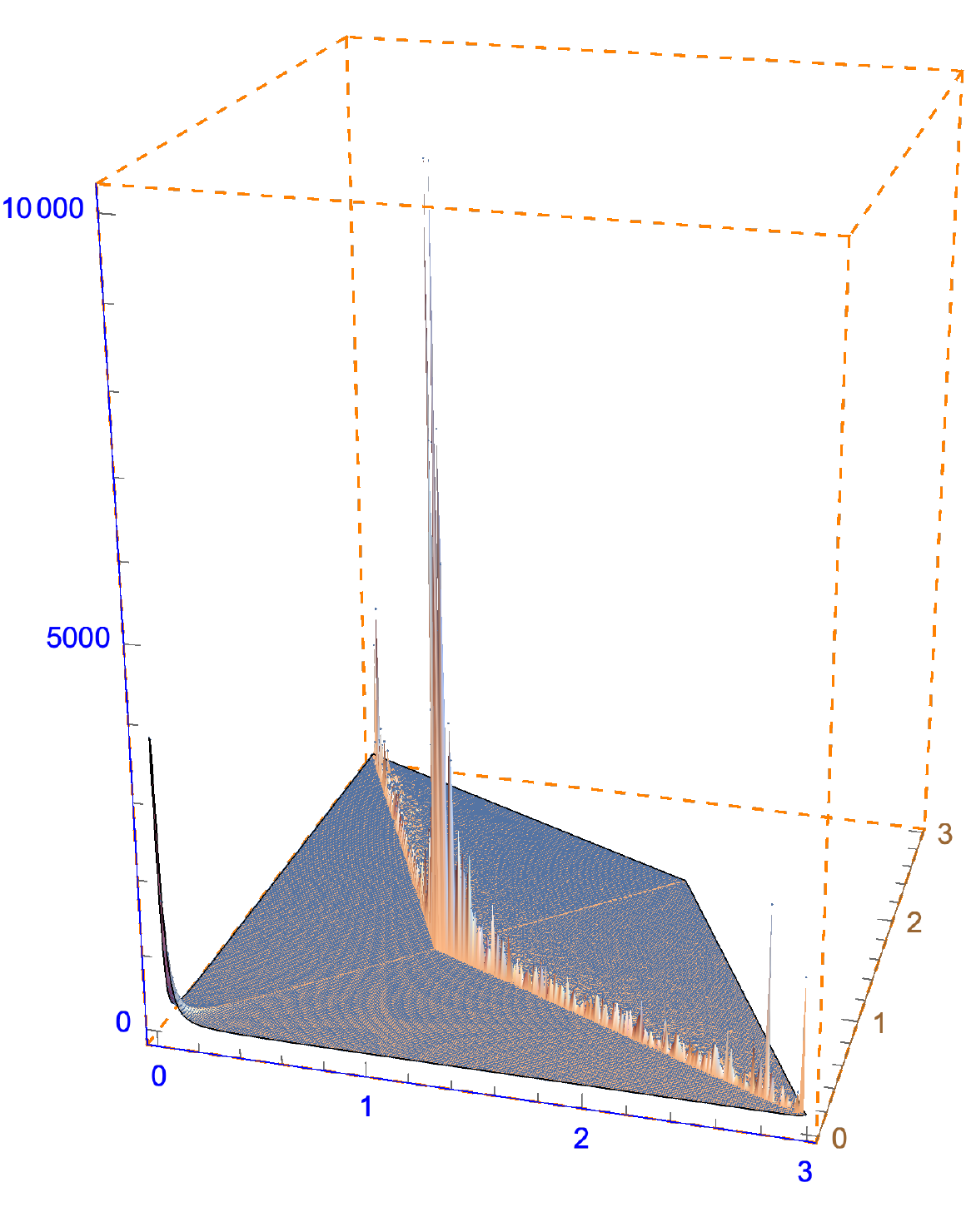}
     \includegraphics[width=10pc]{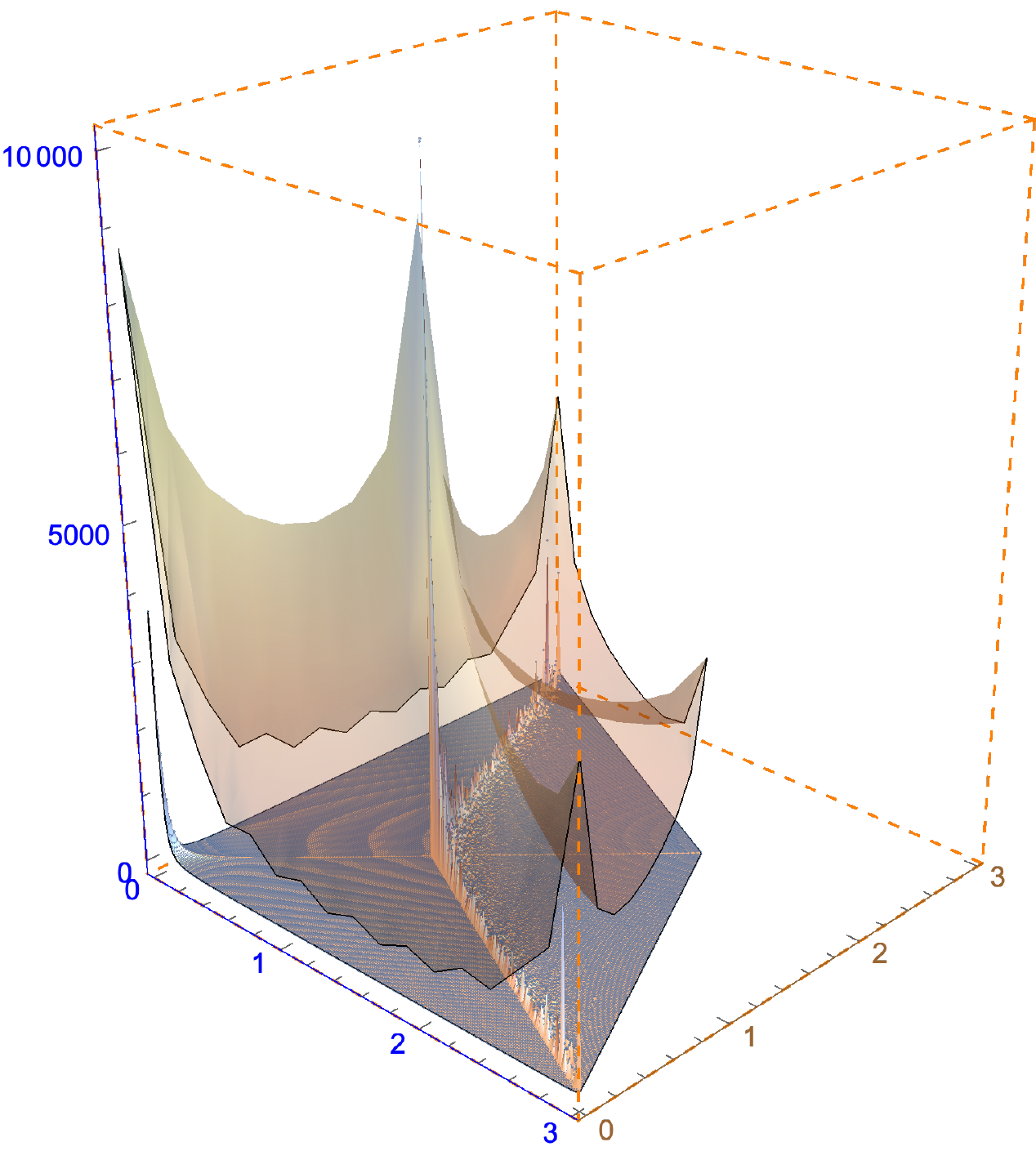}
       \caption{\label{zonalversusJ}  
      \ommit{P a) Semi-transparent surface approximating the vertically rescaled zonal structure constants of $\chi_Z([8,8])^2$.
       b) Point plot of the volume function $J=\rho$ ---  because of the large vertical coordinate range, those parts of the surface lying far from the singularities look essentially flat and one only sees the singularities themselves (high values of $J$).
       c)~Superposition of a) and b).
     The horizontal coordinates (variables $\nu=(\nu_1, \nu_2)$ running between $0$ and $3$) are the Dynkin coordinates corresponding to $\alpha = \beta = \{1,0,-1\}$,  obtained from the partition $\{2,1,0\}$ after a global shift by $-1$.}
     Left: Surface approximating  the rescaled zonal structure constants for some choice of arguments.\\
Middle : Point plot of the volume function $\CJ$ for the corresponding arguments.\\
Right: Superposition of both.\\
Remark : Because of the large vertical coordinate range, those parts of the surface lying far from the singularities look essentially flat and one only sees the singularities (high values of $\CJ$).
}
       \end{figure}
The answer to the first part of the question is not known: notice that there is no clear way to obtain a combinatorial interpretation of a would-be hive or BZ-polytope, since the structure constants of zonal polynomials are usually not integers but rational numbers --- they are integers if the Jack parameter $\theta = 1$, but this is because one recovers in that case the Schur polynomials themselves! We leave this problem to the sagacity of our readers.

 \section*{Acknowledgements} 
The work of Colin McSwiggen is partially supported by the National Science Foundation under Grant No. DMS 1714187, as well as by the Chateaubriand Fellowship of the Embassy of France in the United States.



\newpage
\begin{thebibliography}{99}

\bibitem{BZ} M. Bauer and J.-B. Zuber, {On products of delta distributions and resultants},  to appear


\bibitem{BZ2}  A. Berenstein and A. Zelevinsky, 
Triple multiplicities for $s\ell(r+1)$ and the spectrum of the exterior algebra of the adjoint representation,
{\it J. Alg. Combin.} {\bf 1} (1992) 7--22


\bibitem{BH} \'E. Br\'ezin and S. Hikami, 
An extension of the Harish-Chandra--Itzykson--Zuber integral,
\url{http://arxiv.org/abs/math-ph/0208002v1};
{WKB-expansion of the Harish-Chandra--Itzykson--Zuber integral
for arbitrary $\beta$}, 
{\it Progress of Theoretical Physics}, {\bf 116} (2006)  441--502,
\url{http://arxiv.org/abs/math-ph/0604041v1}




\bibitem{CZ1} R. Coquereaux and J.-B. Zuber, {From orbital measures to Littlewood--Richardson coefficients and
hive polytopes}, {\it Ann. Inst. Henri Poincar\'e Comb. Phys. Interact.}, {\bf 5} (2018) 339--386, 
\url{http://arxiv.org/abs/1706.02793}


\bibitem{CZ2} R. Coquereaux and J.-B. Zuber,  The Horn Problem for Real Symmetric and Quaternionic Self-Dual Matrices,
{\it SIGMA}   {\bf 15} (2019) 029, \url{http://arxiv.org/abs/1809.03394}

\bibitem{CZ3} R. Coquereaux and J.-B. Zuber,  Conjugation properties of tensor product multiplicities, J. Phys. A: Math. Theor. 47 (2014) 455202 (28pp) doi:10.1088/1751-8113/47/45/455202 \url{http://arxiv.org/abs/1405.4887}

\bibitem{CMSZ} R. Coquereaux, C. McSwiggen and J.-B. Zuber,  On  Horn's Problem and its Volume Function, \url{http://arxiv.org/abs/1904.00752}

\bibitem{DST} J. Day, W. So and R. C. Thompson, The spectrum of a Hermitian matrix sum, 
{\it Linear Algebra and its Applications} {\bf 280} (1998) 289--332


\bibitem{Dooley-etal} A.H. Dooley, J. Repka and N. Wildberger, Sums of Adjoint Orbits,  {\it Linear and Multilinear Algebra} (1993),
{\bf 36}, 79--101  



\bibitem{ER} P. Etingof and E. Rains, Mittag--Leffler type sums associated with root systems,
\url{http://arxiv.org/abs/1811.05293}


\bibitem{Fa} J. Faraut, {Horn's problem, and Fourier analysis}, 
Tunisian Journal of Mathematics,
{\bf 1} (2019)  585--606

\bibitem{Fu} W. Fulton, Eigenvalues, invariant factors, highest weights, and Schubert calculus,
{\it Bull. Amer. Math. Soc.} {\bf 37} (2000), 209--249, \url{http://arxiv.org/abs/math/9908012}


\bibitem{HC} Harish-Chandra, {\sl Differential Operators on a Semisimple Algebra},
{\it  Amer. J. Math.} {\bf 79} (1957) 87--120



\bibitem{He82} G.J. Heckman, Projections of Orbits and Asymptotic Behavior of Multiplicities for Compact Connected Lie Groups, {\it Invent. Math.} \textbf{67} (1982), 333--356

\bibitem{Ho54} A. Horn, {\sl Doubly stochastic matrices and the diagonal of a rotation matrix}, {\it Amer. J. Math.} {\bf 76} (1954),
620--630 

\bibitem{Ho62} A. Horn, {\sl Eigenvalues of sums of Hermitian matrices}, {\it Pacific J. Math.} {\bf 12} (1962),
225--241

\bibitem{IZ} C. Itzykson and J.-B. Zuber, {\sl The planar approximation II}, J. Math. Phys. {\bf 21} (1980) 411--421


\bibitem{Kir}  A.A. Kirillov, {\it Lectures on the Orbit Method}.  American Mathematical Society, Providence (2004)



\bibitem{Kly} A.A. Klyachko, {Stable  bundles, representation theory and Hermitian operators}, {\it  Selecta Math.} (N.S.)
{\bf 4} (1998),  419--445

\bibitem{Knut} A. Knutson, The symplectic and algebraic geometry of Horn's problem, {\it Linear Algebra Appl.} {\bf 319} (2000), 61--81

\bibitem{KT00} A. Knutson and T. Tao,
{Honeycombs and sums of Hermitian matrices},  {\it Notices Amer. Math. Soc.} {\bf 48} (2001), 175--186, 
\url{http://arxiv.org/abs/math/0009048}

\bibitem{KT99} A. Knutson, T. Tao, {The honeycomb model of GL(n) tensor products
I: Proof of the saturation conjecture}, {\it  J. Amer. Math. Soc.}
{\bf 12} (1999), 1055--1090,  \url{http://arxiv.org/abs/math/9807160}


\bibitem{Mathematica}Mathematica, Wolfram Research, Inc., Champaign, IL, 2012, \url{http://www.wolfram.com/}


\bibitem{McS} C. McSwiggen, A new proof of Harish-Chandra's integral formula, Commun. Math. Phys. {\bf 365} (2019), 239--253, \url{http://arxiv.org/abs/1712.03995}


\bibitem{AO_Oblades} A. Ocneanu, Various conferences and private communication (2009).

\bibitem{OkounkovOlshanski} A. Okounkov, G. Olshanski, Shifted Jack polynomials, binomial formula, and applications, {\it Math. Res. Lett.} {\bf 4} (1997), 69--78, \url{http://arXiv.org/abs/q-alg/9608020} 




\bibitem{We} H. Weyl, Das \corr{asymptotische} Verteilungsgesetz der Eigenwerte linearer partieller Differentialgleichungen,
{\it Math. Ann.} {\bf 71} (1912), 441--479

\bibitem{Z1}  J.-B. Zuber, {Horn's problem and Harish-Chandra's integrals.  Probability distribution functions}, 
{\it Ann. Inst. Henri Poincar\'e Comb. Phys. Interact.}, {\bf 5} (2018), 309-338, 
\url{http://arxiv.org/abs/1705.01186}





\end{thebibliography}
  \end{document}